\documentclass[twocolumn]{aastex61}

\newcommand{\teff} {\ensuremath{T_{\rm eff}}}
\newcommand{\mic} {\ensuremath{\mu{\rm m}}}

\received{Thermidor 1, II}
\revised{Brumaire 18, IV}
\accepted{\today}
\submitjournal{ApJ}

\shorttitle{The ONC IMF to planetary masses}
\shortauthors{Gennaro \& Robberto}
\usepackage{comment}
\usepackage{amsmath}
\usepackage{multirow}
\begin{document}

\title{HST survey of the Orion Nebula Cluster in the H$_2$O 1.4~\mic~ absorption band:\\
II. The substellar IMF down to planetary masses}

\correspondingauthor{Mario Gennaro}
\email{gennaro@stsci.edu}

\author[0000-0002-5581-2896]{Mario Gennaro}
\altaffiliation{Based on observations made with the NASA/ESA {\it Hubble
Space Telescope}, obtained at the Space Telescope Science Institute, which
is operated by the Association of Universities for Research in Astronomy, 
Inc., under NASA contract NAS 5-26555.  These observations are associated
with program GO-13826.}
\affiliation{Space Telescope Science Institute,
3700 San Martin Drive
Baltimore, 21218, USA}

\author[0000-0002-0786-7307]{Massimo Robberto}
\affiliation{Space Telescope Science Institute,
3700 San Martin Drive
Baltimore, 21218, USA}
\affiliation{Department of Physics and Astronomy,
The Johns Hopkins University, 3400 N. Charles Street, Baltimore, MD 21218, USA}



\begin{abstract}
We exploit the ability of the \textit{Hubble Space Telescope} to probe near infrared water absorption present in the atmosphere of low-mass stars, brown dwarf and planetary mass objects to create a very pure sample of Orion Nebula Cluster (ONC) members, not affected by contamination from background stars and galaxies which lack water absorption. Thanks to these data we infer the Initial Mass Function (IMF) of  
the ONC in the $0.005 - 1.4$M$_{\odot}$ regime, i.e. down to few Jupiter masses.
The young age of the ONC, $\sim1$~Myr, provides a snapshot of the outcome of star formation for the present-day conditions (metallicity, temperature, pressure) of typical Milky Way disk molecular clouds.
We demonstrate that the IMF of the ONC is well described by either a log-normal function or a broken power-law, with parameter values qualitatively in agreement with the canonical Chabrier or Kroupa forms for the Milky Way disk IMF. This continuity in the mass distribution provides clues to the fact that the same physical processes may be regulating formation of stars, brown dwarfs, and planetary mass objects.
Both the canonical IMF forms under-predict the observed number of very low mass members (below 0.1 M$_\odot$), a regime where our data allows more precise constraints.
%Details on the parameter values depend on the assumed star formation history and binary fraction, but they do not affect 
%regardless of the adopted parametrization, the IMF down few Jupiter masses can be described as an extension, with continuity, of the stellar IMF of the Galactic Disk. 
%Even though our best fit models predict a slower decline in number of objects at lower masses with respect to the predictions of the canonical Chabrier/Kroupa IMF models,
Nevertheless, we do not observe a rise or secondary peak in the brown dwarfs or planetary mass regimes. Our study thus contradicts findings based on broad-band near infrared ground-based photometry, which predict an extremely high number of free-floating planets, but likely suffer from unaccounted background contamination. 
\end{abstract}


\section{Introduction}
\label{sec:intro}

One of the most fundamental tests for any theory of star formation is the ability to predict the mass spectrum of the stellar and substellar objects that form within a giant molecular cloud, i.e., the Initial Mass Function (IMF).

Ideal magneto-hydrodynamics and gravity, the physical mechanisms invoked to explain the power-law shape of the IMF in the super-solar mass regime \citep{1955ApJ...121..161S} are scale-free. Therefore other physical mechanisms must be used to explain the observed peak, or characteristic mass, at $\sim 0.2 - 0.3$M$_\odot$  and predict the decay at lower masses \citep{2014PhR...539...49K}. Some plausible mechanisms for explaining the IMF peak and turnover are thermal support \citep{1992MNRAS.256..641L,2005MNRAS.356.1201B}, turbulence \citep{2007ApJ...661..972P,2013MNRAS.433..170H}, and radiative feedback \citep{2009MNRAS.392.1363B,2011ApJ...743..110K}.
In particular \cite{2011ApJ...743..110K} proposes a scenario in which accretion luminosity onto stellar cores can heat the surrounding gas, thus making it more stable against its own collapse. There is a sweet spot in which the accretion luminosity balances the tendency of the cores to collapse, and this corresponds to the peak mass of the IMF.
Early deuterium burning may also play a fundamental role, and this may explain why the typical stellar mass is so close to the hydrogen burning limit \citep{2014PhR...539...49K}.

The formation of brown dwarfs may itself require additional physics  \citep{2007prpl.conf..149B}. Options include dynamical ejection of embryonic brown dwarfs from multiple systems, fragmentation of filaments falling into a cluster potential, or fragmentation of protoplanetary disks, again combined with ejection \citep{2007prpl.conf..459W,2008MNRAS.389.1556B,2009MNRAS.392..413S}. Robust data on the frequency, mass distribution, spatial distribution and kinematics of low-mass stars and brown dwarfs can help clarifying the relevance of each mechanism, but these observational benchmarks remain scarce. 

Young massive clusters (YMCs) are, in principle, ideal laboratories to study the IMF. They are young and have experienced limited stellar and dynamical evolution, thus their observed mass function is as close as possible to be truly initial. They are massive, thus the  IMF is sampled robustly. Their stars constitute a simple stellar population in terms of star formation history and metallicity.
Yet, young massive clusters in our Galaxy present several observational challenges. Being young, they are located in the Galactic Disk behind layers of foreground extinction, and in front of a large column of galactic stellar contaminants. Being rare, they are generally far from us, making it difficult to detect low mass stars; crowding, non-uniform extinction and incompleteness add to the problem of reliably characterizing their stellar and substellar population. 

The Orion Nebula Cluster (ONC) is the ideal candidate for resolved IMF studies. It is young \citep[1-3~Myr,][]{2011MNRAS.418.1948J},
and its moderately large total stellar mass,  $\sim 2500 M_{\odot}$\citep[estimate from][using a density profile fit to the observed counts of stellar and substellar objects up to 2 pc from the ONC core]{2014ApJ...795...55D}  allows for a robust sampling of the IMF. 
Its proximity, $403^{+7}_{-6}$~pc \citep{2019ApJ...870...32K}, allows reaching the brown dwarf regime.
The ONC size on the sky (order of half a degree radius) and location toward the galactic anticenter, $(l,b) =(209.0085 \,\mathrm{deg}, -19.3828 \,\mathrm{deg})$ for the Trapezium\footnote{as reported by \url{http://simbad.u-strasbg.fr}}, imply that crowding is much reduced with respect to more compact and distant young massive clusters typically located in the two inner galactic quadrants.
For all these reasons, the ONC has been the target of many IMF studies \citep[see e.g.][]{2000ApJ...540..236H,2002ApJ...573..366M,2012AJ....144..176D} providing a key benchmark for comparison with the IMF measured in different star-forming regions, in young open clusters, as well as in the galactic field. 

\begin{figure*}[!t]
\begin{center}
\includegraphics[width=\textwidth]{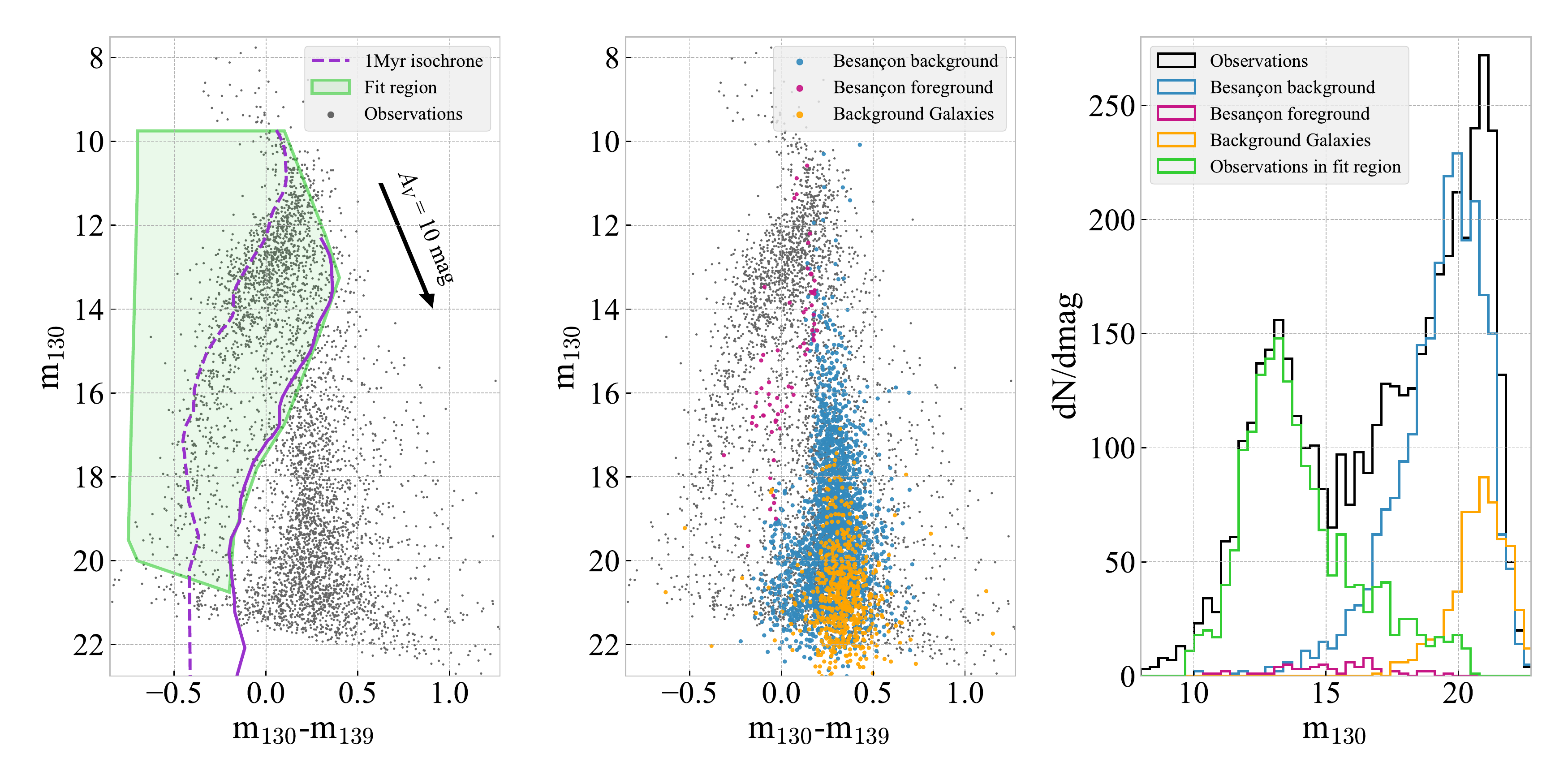}
\caption{Left: the observed color-magnitude diagram. The green shaded area highlights the region used for the IMF fit. The faint limit of such region is chosen to be above the magnitudes where incompleteness becomes an issue. The red limit is chosen to approximate the location of a 1 Myr isochrone, at the average distance of the ONC \citep[403 pc, or $\mu$ = 8.03][]{2019ApJ...870...32K}, reddened by $A_V = 10$ mag,  i.e. the maximum $A_V$ value expected for ONC members, according to \cite{2011AA...533A..38S} (see also Figure~\ref{fig:SpDistr} for details on the $A_V$ distribution). The 1 Myr isochrones are shown in purple, with the dashed and solid lines representing the $A_V = 0, 10$ mag cases respectively. The direction of the reddening vector when using an $R_V =3.1$ \cite{1989ApJ...345..245C} extinction law is shown by the black arrow, with magnitude corresponding to $A_V = 10$ mag. Center: location of simulated contaminants overlaid on the data. Right: luminosity functions in the F130N band for the observed data, the observed data within the green shaded region of the left panel, and the predictions for the contaminating populations. This Figure summarizes the contents of Figures~7, 8 and 15 of Paper~I, with some minor modifications.
\label{fig:CMDs} } 
\end{center}
\end{figure*}

Of particular interest in the recent past is the topic of how the IMF of the ONC extends into the brown dwarfs regime.
The stellar IMF in the Milky Way disk is well characterized has having a characteristic mass, or peak above the hydrogen burning limit, at 0.2-0.3 M$_\odot$ \citep[see e.g.][]{2010AJ....139.2679B} and a decline at lower masses. This qualitative understanding is independent on the parametrization chosen to describe the IMF, either a log-normal \citep{2003PASP..115..763C} or a broken power law \citep{2001MNRAS.322..231K}.
%Studies of the IMF down to the hydrogen burning limit consistently indicate a decline at lower masses. This qualitative understanding is independent on the parametrization chosen to describe the IMF, either a log-normal \citep{2003PASP..115..763C} or a broken power law \citep{2001MNRAS.322..231K}.
%in the Galactic field, are %however 
%hampered by the faintness of old brown dwarfs.
The uncertainties increase as one moves to  masses below the deuterium burning limit, in the planetary mass regime ($\sim 3$ to 15 M$_{Jup}$), as only a handful of candidates have been identified in young Galactic star clusters \citep{2008A&A...477..895Z,2010ApJ...709L.158M}.
The debate about the shape of the substellar IMF, i.e. if it generally follows - and how closely - a Kroupa/Chabrier-type IMF, has been recently enriched by  \cite{2016MNRAS.461.1734D} who suggest that the substellar IMF of the ONC is  bimodal with a peak at about 0.25~M$_\odot$, a  pronounced dip at the hydrogen burning limit (0.08M~M$_\odot$) and a second peak at 0.025~M$_\odot$. 
%The current study uses a different data set and a different technique to test for the presence of such secondary IMF peak below the H-burning limits.

In this paper we investigate the shape of the IMF of the ONC in the low stellar mass regime and down to the the sub-stellar and planetary mass regimes.
We use the near infrared data obtained by a 52 orbits Hubble Treasury Program (GO-13826, PI: M. Robberto). The survey strategy and data analysis has been presented in the first paper of this series (Robberto et al. 2020, hereafter Paper~I) and will be briefly summarized in Section~\ref{sec:data}.
Section~\ref{sec:method} describes the method used to simulate synthetic CMDs, and the techniques we adopt to compare simulations and observations and derive a probability distribution for the IMF parameters. 
In Section~\ref{sec:res} we describe the main results and we discuss them in detail in Section~\ref{sec:discuss}. We summarize our findings and conclude in Section~\ref{sec:sumandconc}.

%%%%%
%%%%%
%%%%%
%%%%%

\section{The data}
\label{sec:data}
%%%%%
%%%%%
%%%%%
%%%%%

%%%%%%%%%%%%%%%%%%%%%%%%%%%%%%%%%%%%%%%%%%%%%%%%%%%%%%%%%%%%
%%%%%%%%%%%%%%%%%%%%%
%%%%%%%%%%%%%%%%%%%%%%%%%
%%%%%%%%%%%%%%%%%%%%%%%%%
%%%%%%%%%%%%%%%%%%%%%
%%%%%%%%%%%%%%%%%%%%%%%%%%%%%%%%%%%%%%%%%%%%%%%%%%%%%%%%%%%%%%%%%%%%%%

The WFC3/IR imaging data considered in this paper were taken using the F130N and F139M filters, chosen to sample the 1.4$\mu m$ water absorption band (F139M) and the adjacent line-free continuum (F130N). Water vapor is known to be present in the atmospheres of cool stars, but its most prominent spectral signatures can only be detected at high significance from  space, unobstructed by telluric water vapor absorption.
The depth of the 1.4~$\mu$m absorption feature becomes more and more pronounced as the temperature of low-mass stars decreases (see Paper~I, Figure~9). The $m_{130} - m_{139}$ color becomes correspondingly more negative, i.e. bluer. The ONC data, however, show that this trend ends at colors $m_{130} - m_{139} \sim - 0.35$ mag, or $T_{\mathrm{eff}} \sim 2700$K for a 1~Myr isochrone. Below this temperature, saturation occurs and the $m_{130} - m_{139}$ color remains constant at $\sim - 0.35$ mag down to planetary masses (see Figure~\ref{fig:CMDs}, left).

The characteristic locus of low-mass stars and sub-stellar objects in a $(m_{130} - m_{139}, m_{130})$ color-magnitude diagram (CMD) allows to discriminate cluster members from  background, reddened stars and galaxies that do not show such absorption, confirming that this combination of filters provides a very effective means of identifying the low-mass population of the ONC. Such separation is not possible using broad-band near infrared filters.

We refer the reader to Paper~I for a detailed discussion of the observations and data reduction steps. In Paper~I we describe the procedure for generating the artificial star tests as well. The latter are adopted in this paper to produce synthetic CMDs that are compared to the data to obtain the best-fit IMF.

Paper~I also details a procedure to account for foreground as well as background (galactic and extra-galactic) contamination. Figure~\ref{fig:CMDs}, center, shows the CMD position of the expected contaminants. The Galactic contamination is estimated using the Besan{\c{c}}on model of the Milky Way \citep{2003A&A...409..523R}. The extragalactic contribution is obtained by computing synthetic photometry from best-fit spectra to CANDELS galaxies \citep{2012MNRAS.421.2002P}. In both cases the contaminants synthetic photometry accounts for the surveyed area and for the extinction distribution towards our field of view, as provided by \cite{2011AA...533A..38S}. More details on contamination and the determination of a pure sample of ONC members are given in Section~\ref{sec:CMFFR}.

\section{Deriving the IMF}
\label{sec:method}

Our goal is derive a posterior distribution for a set of parameters that describe the IMF.
Examples of such parameters are the slopes and break points for a broken power law or the peak mass and width of a log-normal distribution.
Our goal is achieved by first simulating synthetic CMDs based on the IMF parameters values and then by comparing such simulations to the observations. We then iterate the process using MCMC techniques to obtain a probability distribution for the parameters of interest.

Our methodology can be separated in two broad aspects:
\begin{itemize}
\item CMD simulations: this aspect entails providing an accurate description of the observed data in terms of (i) the underlying physical properties of individual stars (mass, age, chemical composition, binary propoerties), (ii) a membership criterion to separate ONC members and background contaminants, (iii) a model of the spatial distribution of stars in the cluster and of their extinction, (iv) the observational uncertainties. 
The process of simulating CMDs and the underlying ingredients are described in Section~\ref{sec:sumCMD}.
\item Fitting: this aspect consists in defining a reasonable parametrization for the IMF, defining a rigorous procedure to compare the simulations and the observations, and derive a probability distribution function
for the IMF parameters. The fitting technique is described in Section~\ref{sec:fit}.
\end{itemize}

\begin{figure*}[t]
\begin{center}
\includegraphics[width=0.98\textwidth,trim={0cm 2cm 0 2cm},clip]{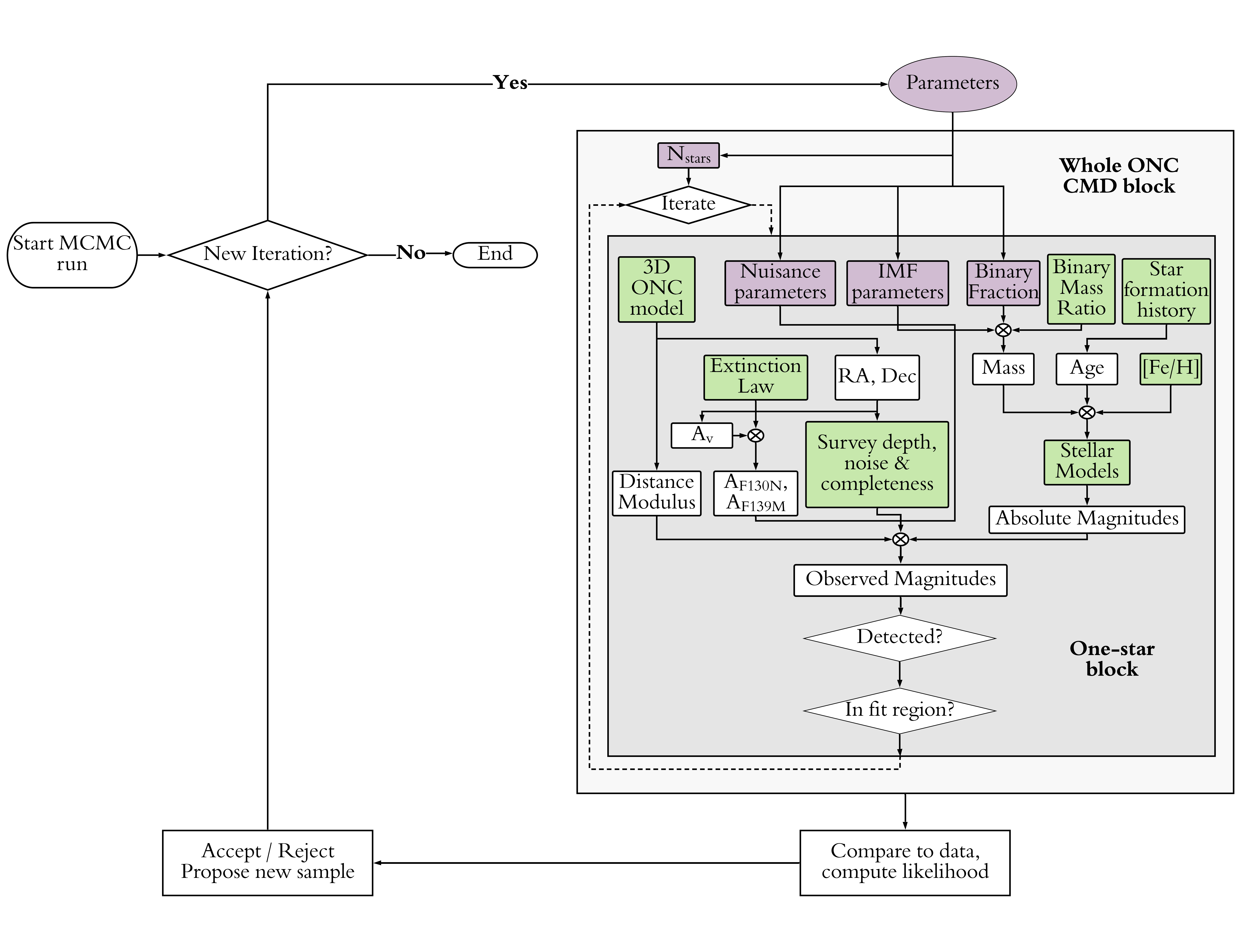}
\caption{Flowchart for the CMD simulations and IMF fitting procedure. The section in dark gray represents the simulation of a single star into the CMD. The lighter gray section surrounding the darker one represents the simulation of the entire CMD of ONC members. The purple colored blocks are the MCMC fitting parameters. The green colored blocks are the fixed ingredients of the CMD simulations.\label{fig:lucid}} 
\end{center}
\end{figure*}

The overall methodology is illustrated in Figure~\ref{fig:lucid}.

\subsection{Simulating the ONC CMD}
\label{sec:sumCMD}

The following steps are necessary to simulate a single star in the CMD:
\begin{enumerate}
\item We draw a total mass using the IMF parameters and we draw a probability of the star being a binary using the binary fraction paramters.  IMF parameters and binary fraction paramter are iterated upon in the fitting procedure described in Section~\ref{sec:fit}. We also draw a binary mass ratio and an age from fixed distributions and assumed solar metallicity. This step is detailed in Section~\ref{sec:stpar}. 
\item We use stellar models to obtain absolute magnitudes, $M_{130}$ and $M_{139}$, based on the extracted parameters. This step is detailed in Section~\ref{sec:models}. 
\item We use a 3D model of the cluster to simulate the stellar position in the ONC. From this we derive a distance modulus, $DM$. This step is detailed in Section~\ref{sec:spdistr}. 
\item Using the stellar location and a 2D extinction map, we determine $A_V$. From this we compute $(A_{130}, A_{139})$ using the \cite{1989ApJ...345..245C} extinction law. This step is detailed in Section~\ref{sec:extmap}
\item At this stage we thus have a noiseless set of magnitudes $(m_{130}, m_{139})_{intr} = (M_{130}, M_{139}) + (A_{130}, A_{139}) + DM$. The next step is to use artificial star tests to determine whether the simulated star can be detected as well as its observed, noisy magnitudes $(m_{130}, m_{139})_{noisy}$. This step is detailed \ref{sec:ASstars}.
\item Based on simulations of the background contaminants and consideration on where the ONC members are located in the CMD we define a CMD fitting region where we expect most of the members will fall, without significant contamination.
The simulated noisy magnitudes are used to test whether the star falls within such CMD region and the star is either kept or rejected accordingly. This step is detailed in Section~\ref{sec:CMFFR}.
\item Based on the total true number of stars (a variable in our MCMC fitting procedure), we iterate all the above steps. The number of stars that land in the fitting region is smaller than the true number of stars, both because of incompleteness and because of our membership selection criterion. This number, with uncertainty, tends toward the number of observed stars in the fitting region. This step is detailed in Section~\ref{sec:totnum}.
\end{enumerate}

\subsubsection{Stellar parameters}
\label{sec:stpar}
In order to simulate a star's magnitudes we need to know its mass, whether it is a binary, its age and its metallicity.
The total system mass is drawn from an IMF, whose parameters are what we are fitting for.
We then use the binary fraction, which is a fit parameter, to determine whether the system should be single or a binary.
This is done by drawing a random number from a uniform distribution between 0 and 1, and if the number is smaller than the binary fraction, we consider the star to be a binary, we consider it a single star otherwise.
In the case of binaries, given the system total mass and the binary mass ratio, we determine the mass of each component and sum their fluxes.
The mass ratios themselves are randomly drawn from a uniform distribution between 0 and 1, a valid assumption for solar type stars, which we extrapolate to lower masses \citep{1991A&A...248..485D,2010ApJS..190....1R}.

The stellar age is drawn from a fixed star formation history. For each fitting run we use either one of: 1) a single episode of star formation, 1 Myr ago, or 2) a uniform star formation rate between 1 and 3 Myr ago (i.e. a constant number of stars formed per unit time along such interval). As we show in Section~\ref{sec:res}, some details in the results depend on the choice of the star formation history.
An assumption of our work is that stellar age and stellar mass are independent variables. This is equivalent to saying that at any age under consideration in this work, 1-3 Myr, the \textit{instantaneous} mass distribution of newly formed stars is the same -- i.e. the IMF is time-invariant. 

Moreover we postulate that no dynamical process has yet altered the shape of the observable present-day mass function with respect to the IMF. This is equivalent to saying that if stars born in the ONC have been lost to the Galactic field, such loss is mass-independent.
To corroborate this assumption we proceed thus: considering a velocity dispersion of a few km s$^{-1}$ and a cluster size of a few pc, we have that the crossing time for the ONC cannot be shorter than $t_{cross} = 0.1$~Myr. Using the classical formula by \cite{1969ApJ...158L.139S}, $t_{relax} \approx N/(8\log N) \times t_{cross}$, where $N\sim 2000$ is the number of ONC members, we obtain $t_{relax} \sim 3.3 $~Myr. The relaxation time regulates the energy equipartition in the system, thus the fact that the ONC is younger than its relaxation time, ensures that the kinetic energy distribution is not yet in equilibrium. This in turn implies that any mass loss the ONC might have experienced (i.e. stars that are ejected outside our surveyed field of view) must have been "gray", i.e. the ONC should not have preferentially lost low mass members (faster if in equipartition) with respect to high mass ones (slower). 
The assumption of non-equipartition is corroborated by the recent study by \cite{2017ApJ...845..105D} who find that although in global virial equilibrium, or possibly slightly supervirial state, the ONC \textit{individual} velocities do not depend on the individual masses, a sign that equipartition has not been reached yet.
This justifies our assumption that the mass and age distributions can be treated independently, as well as that the measured mass function is truly -at a minimum proportional to- the \textit{initial}.

The metallicity is held fixed at the solar value, $\mathrm{[Fe/H]} = 0.0$ dex. This is a very good assumption for the ONC, for example \cite{2009A&A...501..973D} find a metallicity of  [Fe/H]~$=-0.01 \pm 0.04$~dex for the low mass stars in the ONC.

Given masses (for either a single object or the two components of a binary), age, and metallicity the absolute $M_{130}$ and $M_{139}$~magnitudes are computed using the models described in Section~\ref{sec:models}.

\subsubsection{The stellar models}
\label{sec:models}

We utilize the stellar models by \cite{2015A&A...577A..42B} in the version available on the PHOENIX-Lyon website\footnote{\url{http://perso.ens-lyon.fr/france.allard/}}. These models reach down to 1~M$_\mathrm{Jup}$. In particular, we adopt the BT-Settl version of the model atmospheres by \cite{2012RSPTA.370.2765A} with  solar metallicity and solar-scaled abundances according to \cite{2009ARA&A..47..481A}. We concentrate on the isochrones in the 1 to 3 Myr age range.

Paper~I shows that these models are not able to reproduce the $m_{130}-m_{F139}$ colors characteristic of the low-mass ONC population below \teff $\sim 2700$K. A semi-empirical color correction to the models has been introduced in Paper~I. In the current work we adopt the corrected models and use them to assign synthetic absolute $M_{130}$ and $M_{139}$ magnitudes to simulated stars.

%A 1 Myr isochrone is overlaid to the data in Figure~\ref{fig:CMDs}.
\subsubsection{The spatial distribution}
\label{sec:spdistr}
\begin{figure*}[t]
\begin{center}
\includegraphics[width=0.98\textwidth]{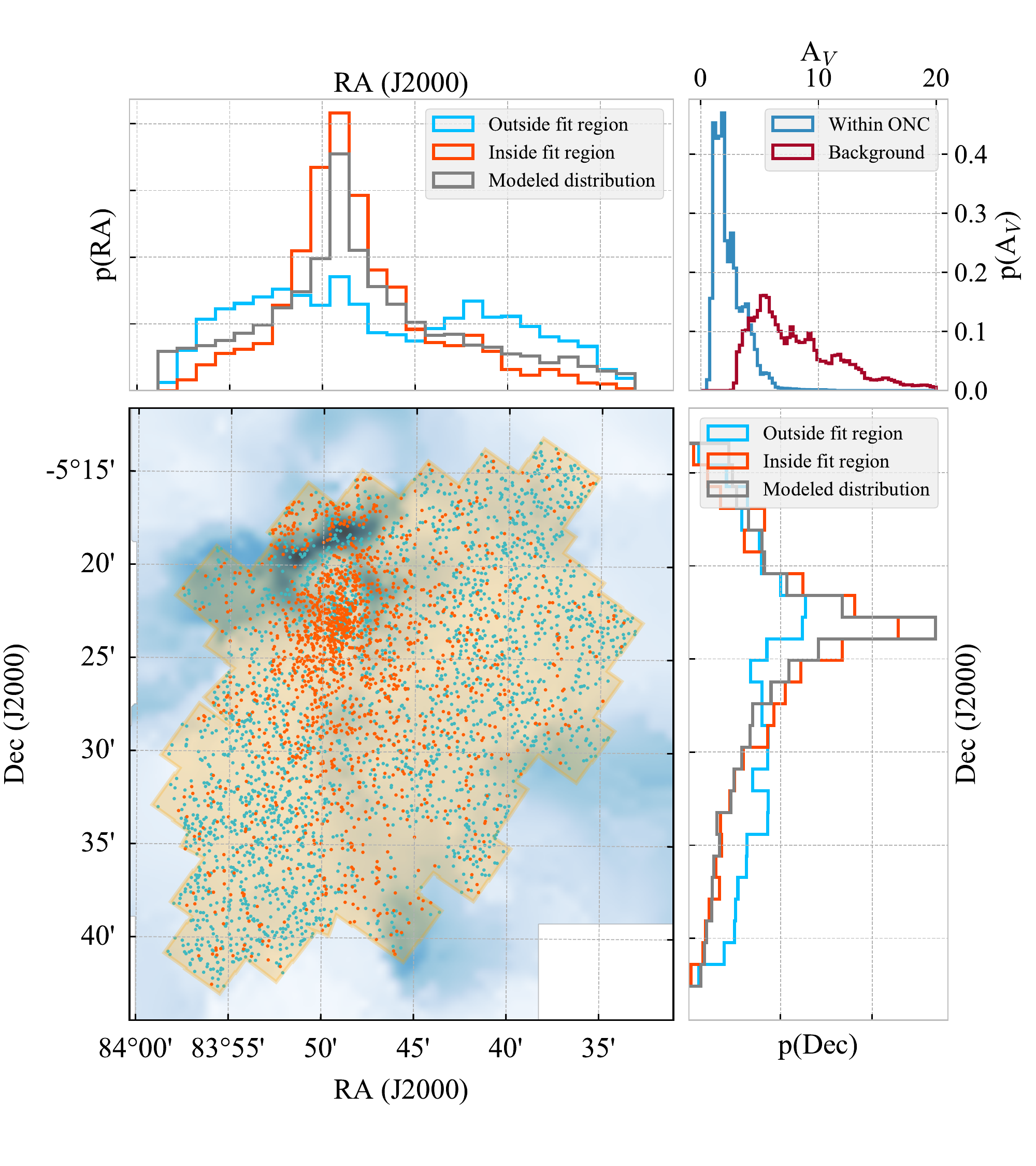}
\caption{Spatial distribution of all the observed sources, with the sources within the fit region (green shaded area in Figure~\ref{fig:CMDs}, left panel) in orange, and the other sources in cyan. The orange shaded area  in the main panel corresponds to the current WFC3/IR observations footprint. The upper left and lower right panels show the marginal RA and Dec distributions of the data. Overlaid in brown in such panels is the histogram for 2000 objects drawn from the adopted model spatial distribution, a spherically symmetric distribution with $\rho \propto \rho_0 r^{-0.2}$ and center in (RA, Dec) = (05:35:16.26 hours, -05:23:16.4 degrees), and landing within the observations footprint. The background image in the main panel is the foreground extinction map adopted for our simulations of ONC members \citep{2011AA...533A..38S}. The map used for estimating the extinction of simulated background sources (Figure~\ref{fig:CMDs}) is not shown \citep[such map is also from][]{2011AA...533A..38S}. The upper right panel shows the histogram of $A_V$ values for both the foreground extinction map used for ONC objects (blue) and the foreground plus ONC molecular cloud extinction map used background objects (red). \label{fig:SpDistr}} 
\end{center}
\end{figure*}

As visible in Figure~\ref{fig:SpDistr}, the objects within the CMD fitting region (orange) have a centrally concentrated distribution, while the objects outside the same region (cyan) are much more uniformly distributed.
The definition of the CMD fitting region is provided in Sect.~\ref{sec:CMFFR}, it suffices to say here that this is the CMD region where background contamination is minimal and the cool, low-mass ONC members are bluer than the background objects and clearly separated from them.

The detectability of a source of given intrinsic magnitude depends on the local crowding around that source, on the brightness of the local background, and on the extinction in front of such source. All these factors are highly spatially variable in the ONC. 
In simulating a CMD, the assumed position of a source contributes to determining whether it will 
end up being detected and where it will lie in the CMD.
This means that one has to assume an underlying 3D spatial distribution, aiming at producing not only the best match to the observed CMD, but also an observed spatial distribution in good agreement with the data.

In order to determine a good representation of the true 3D distribution of ONC members, 
we assume no spatial dependence of the IMF, in the sense that $p(m, \mathrm{RA},\mathrm{Dec}) = p(m)*p(\mathrm{RA},\mathrm{Dec})$, where $p$ are probability distribution functions, $m$ is the stellar mass, and $\mathrm{RA}$, $\mathrm{Dec}$ are the on-sky coordinates of the ONC members.
We in fact verify this assumption \textit{a posteriori}, see below in this Section.

We further assume that $p(\mathrm{RA},\mathrm{Dec})_{intr}$, i.e. the intrinsic distribution on the sky, can be derived from a 3D distribution with  
spherical symmetry, centered on the ONC center, and that such distribution can be parametrized as a power law. We thus have $p(\mathbf{r}) =  p(r,\theta,\phi) = \rho(r)  = \rho_0 * r^{-\gamma} $, where $|\mathbf{r}| = r$. \cite{2014ApJ...795...55D} found an index $\gamma = 2.4$ for ONC members with X-ray counterparts. Such an index denotes a very steep, centrally concentrated distribution for the ONC stars in the central cluster.
The $\alpha$ value chosen by \cite{2014ApJ...795...55D} gives however a poor fit for our population of ONC members, which includes objects of much lower mass. The fact that our very low mass stars are much less concentrated, or more uniformly distributed, than the more massive stellar sample of \cite{2014ApJ...795...55D} extends towards the lowest mass range the well known tendency of the ONC to have an overabundance of massive stars at the center of the cluster, i.e. to be mass-segregated. \citep[see e.g.][]{1998ApJ...492..540H,2011MNRAS.415.1967A}.

We adopted the same center of \cite{2014ApJ...795...55D} for our distributions: (Ra, Dec)  = (05:35:16.26, -05:23:16.4).
The assumed distance for the ONC (which matters when projecting on the sky our 3D distribution) is 403~pc, as reported by \cite{2019ApJ...870...32K} based on GAIA DR2 data \citep{2018A&A...616A...1G}.
Note that the 403~pc distance is in slight disagreement with the VLBI measurements of 388~pc by \citep{2017ApJ...834..142K}, who later refined their distance including also GAIA DR2 constraints to $386\pm3$~pc \citep{2018AJ....156...84K}.
\cite{2019ApJ...870...32K} explain the differences in terms of the different survey regions used by \cite{2017ApJ...834..142K,2018AJ....156...84K} and their own work. The \cite{2019ApJ...870...32K} work focuses the ONC central region, while \cite{2017ApJ...834..142K,2018AJ....156...84K} utilize a larger area.
Given the proximity of the ONC and its three dimensional structure, using different regions corresponds to probing different depths into the Orion Complex.
According to \cite{2019ApJ...870...32K}, the central cluster is embedded a dozen parsecs behind the average location of the more distributed YSOs measured in non-thermal emission by \cite{2017ApJ...834..142K}.
Nevertheless, part of the discrepancy might as well be due to a systematic zero-point offset between Gaia DR2 measurements ad VLBI ones, as reported by \cite{2018AJ....156...84K}.

\begin{comment}
Its proximity, 388~pc \citep{2017ApJ...834..142K}, allows reaching the brown dwarf regime. Note that we adopt the 388~pc distance, based on VLBI measurements, throughout this work. \cite{2018AJ....156...84K} refined their ONC distance estimate to $386\pm3$~pc using VLBI and GAIA DR2 data \citep{2018A&A...616A...1G}. \cite{2019ApJ...870...32K} find a distance of $403^{+7}_{-6}$~pc using only GAIA DR2 data. The latter paper explains the differences in terms of the different survey regions used by \cite{2018AJ....156...84K} and their own work. Given the proximity of the ONC and its three dimensional structure, using different regions corresponds to probing different depths into the Orion Complex. Part of the discrepancy might as well be due to a systematic zero-point offset between Gaia DR2 measurements ad VLBI ones, as reported by \cite{2018AJ....156...84K}.
\end{comment}

We did not perform a formal fit to determine the best value of the density distribution slope, $\gamma$. Rather, we assumed a \cite{2001MNRAS.322..231K} IMF, and simply swept a range of values for $\gamma$. We simulated CMDs using the whole technique described in Sections~\ref{sec:stpar}--\ref{sec:totnum}, thus obtaining the observed on-sky distribution $p(\mathrm{RA},\mathrm{Dec})_{obs}$.
We then plotted the histograms for RA and Dec separately, 
and compared those to the histograms for the observed data.
Our preferred slope for the very low mass object spatial distribution is  $\gamma = 0.2$, much shallower than the stellar value of \cite{2014ApJ...795...55D}.
Simulations using this value are overplotted in the histrograms of Figure~\ref{fig:SpDistr}. 

We verified that changing the input IMF slope does not change the output 2D on-sky distribution significantly, and thus our assumption that IMF and 3D distribution can be treated as independent holds, at least for practical purposes.

It is worth noting that our procedure does not account for possible ellipticity of the low-mass stars distribution \citep[see again][]{2014ApJ...795...55D}, but as Figure~\ref{fig:SpDistr} shows, the modeled distribution reproduces the observed one rather well for such a simple parametrization.

In our synthetic CMD generation procedure, the 3D position of each star is drawn from the $\rho_0 * r^{-\gamma} $ distribution above, independent on the other stellar parameters. The RA and Dec are used to compute the line-of-sight extinction (Section~\ref{sec:extmap}) as well as spatial dependent completeness and photometric errors (Section~\ref{sec:ASstars}). The line-of-sight distance from the 3D position is converted into a distance modulus that is added to the absolute magnitudes.

\subsubsection{The 2D extinction map}
\label{sec:extmap}
The ONC is still partially embedded in its parental molecular cloud, rich in gas and dust and with a highly complex 3D structure that causes a high degree of variability in the line-of-sight extinction.
\cite{2011AA...533A..38S} analyzed the extinction toward the ONC using infrared photometry, deriving two separate extinction maps for the two main groups of sources: ONC bona-fide members and background sources. The first group is affected by ``foreground'' extinction, expected to be highly irregular since ONC sources that appear close to each other in projection may lie at different depth within the molecular cloud. On the contrary the background sources, lying behind the whole Orion molecular cloud, typically have higher extinction but smoother spatial distribution of $A_V$ values.
The \cite{2011AA...533A..38S} ``foreground'' extinction map for the ONC is shown in Figure~\ref{fig:SpDistr}. The top-right panel of the same figure shows the histograms of $A_V$ values for both the ONC and the background populations.

In Paper~I we assessed the level of contamination from background sources in the CMD. For those sources,  we assign to each a value of the extinction randomly extracted from the Orion Molecular Cloud extinction map. The results are reproduced in the center panel of Figure~\ref{fig:CMDs}, and show how the simulated background stars and galaxies all lie in the $m_{130}-m_{139} \sim 0$ region, well separated from the blue population of ONC members (Section~\ref{sec:CMFFR}).

For simulating the CMD of ONC member stars, we use the foreground extinction map  as a starting point.
Given a simulated 3D position and corresponding RA, Dec (Section~\ref{sec:spdistr}), we use the extinction map value at that position to assign a mean extinction. The \cite{2011AA...533A..38S} map has a typical resolution of few arcminutes and therefore it provides a good average value of $A_V$.
However, due to the high spatial variation in the intervening gas structure, we
expect the actual value of $A_V$ for a star in that position not to be exactly equal to the mean value from the map.
We use the fact that the distribution of $A_V$ values is induced by the turbulent nature of the ISM, which can be well described by a log-normal function. Typical values of the width of the log-normal are in the $\sigma_{\mathrm{ISM}} = 0.15 - 0.30$ range \citep{2017NJPh...19f5003K}. Thus, to account for the variability of $A_V$ we use $A_{V, \mathrm{map}}$ and fix $\sigma_{\mathrm{ISM}}$ to $0.30$ to define a log-normal distribution with $(\mu, \sigma)$ = $(A_{V, \mathrm{map}}-0.5\sigma_{\mathrm{ISM}}^2, \sigma_{\mathrm{ISM}})$.
A log-normal defined this way has mean equal to $A_{V,\mathrm{map}}$.
The actual extinction value assigned to a star simulated at a given position is a random draw from the log-normal$(\mu, \sigma)$. Varying $\sigma_{\mathrm{ISM}}$ within the $0.15-0.30$ range does not affect the final results for the IMF parameters.

Given $A_V$ we use the \cite{1989ApJ...345..245C} extinction law to compute $(A_{130}, A_{139})$. Since we are using narrow and medium band filters we assume the ratios of extinction values in the $V$ band and the $F130N$, $F139M$ bands to be independent on spectral-type.
We compute the ratios at the bandpasses effective wavelength, $\lambda_{eff, F130N} = 13006 $\AA,  $\lambda_{eff, F139M}= 13838$ \AA~to be $A_{130}/A_V = 0.264$ and $A_{139}/A_V = 0.241$. 

Adding together the absolute magnitudes, the distance modulus, and the extinction produces "intrinsic", i.e. noiseless apparent magnitudes, $(m_{130}, m_{139})_{intr}$.

\subsubsection{Survey area coverage and artificial star tests}
\label{sec:ASstars}
The $(m_{130}, m_{139})_{intr}$ noiseless values are converted into noisy magnitudes using artificial stars experiments.

The ONC survey strategy is amply detailed in Paper~I. We note here that due to the adopted dither pattern, the coverage of the ONC area is somewhat irregular, with regions visited anywhere between 1 and 4 times. The strategy for producing the data catalog consists in averaging together the individual values of the photometry, when the same star is imaged in multiple exposures.
The survey depth and photometric errors are thus dependent on both the local crowding and diffuse background values, as well as on the number of repeated observations available.

In generating the synthetic CMDs used for IMF fitting, we reproduce the same strategy adopted to extract the real source catalog.
We start from a randomly extracted 3D stellar position (Section~\ref{sec:spdistr}), projected on the sky, to create a simulated {\sl model} star of known magnitude. 
We then determine the individual exposures $j$ where the model star would fall. Since
the artificial stars experiment, discussed in more details in Paper I, has been conducted at the exposure level, 
we use those results to determine whether the model star is detected in each individual exposure, with its measured magnitude.
Specifically, for each exposure $j$ we proceed as follows: we select all artificial stars used to run the completeness test within 0.1 mag of $(m_{130}, m_{139})_{intr}$ of our model star, using a 2D euclidean distance in magnitude-magnitude space.
A search radius of 0.1 mag ensures that the completeness is not changing significantly within this interval, while guaranteeing that several tens of artificial star tests are available. 
We then pick the artificial star closest on the sky to the model star.
If the selected artificial star is not recovered in either one of the photometric bands, neither is the model star, otherwise the model star is detected with noisy magnitudes
$(m_{130}, m_{139})_{noisy, j} = (m_{130}, m_{139})_{intr} - (m_{130}, m_{139})_{AS, input} + (m_{130}, m_{139})_{AS,output}$, where $AS$ indicates the chosen artificial star. This formula accounts for the small offset in magnitude between the intrinsic model value and the input magnitude of  the closest artificial star.
If the on-sky position of the model star is such that our survey covers that position with $j>1$ individual exposures, the individual $(m_{130}, m_{139})_{noisy, j}$ measurements are averaged over the exposures, only for the cases in which the output artificial star is recovered.
If the artificial star is never recovered in the $j$ exposures, then the model star is considered a non-detection. Non-detections still count towards the total number of generated stars (Sect.~\ref{sec:totnum}).

\subsubsection{The ONC members sample}
\label{sec:CMFFR}

We identify a region of the CMD where the population can be considered as purely consisting of ONC members, with little or no contribution from fore- and background contaminants. The region is highlighted by the green shaded area in the left panel of Figure~\ref{fig:CMDs}. 

To define this region, which we also refer to as the fit region, we use the fact that the maximum expected extinction for ONC objects in the surveyed area is no larger than 10 magnitudes in $A_V$, as visible in Figure~\ref{fig:SpDistr}.
A 1~Myr isochrone is shown in the left panel of Figure~\ref{fig:CMDs} at the average distance of the ONC \citep[388 pc,][]{2017ApJ...834..142K} and with $A_V$ equal to either 0 or 10 mag (the average extinction is of the order of $A_V=2.5$ mag). The $A_V = 10$ mag isochrone is used to delimit the reddest edge at which we expect ONC members. This limit cuts off the greatest part of the expected contaminants as well. 
On the faint end, we adopt a cut corresponding to $m_{139} = 21$ mag, the average 50\% completeness limit throughout the field of view (local variations due to crowding and background level are discussed in Paper~I); the adopted limit can be see as a slanted line, parallel to the detection limit in the CMD of Figure~\ref{fig:CMDs}, left panel (the vertical axis in the CMD is $m_{130}$, not $m_{139}$, hence the faint limit line is not horizontal). The adopted magnitude limit corresponds in mass to 0.005 - 0.007 M$_{\odot}$, about $5-7$ M$_\mathrm{Jup}$.
The bright limit of the fitting region corresponds to the location of a 1.4 $M_{\odot}$ star, the most massive in the adopted models. We follow this point along the reddening vector to define the bright, red end of the fitting region corresponding to diagonal line going from about (0.1,10) to about (0.4, 13) in the CMD.
We do not go through the exercise of patching the isochrone with one including higher mass stars, because the data themselves start showing signs of saturation above $m_{130} = 10$ mag.
The blue limit of the fitting region is a simple straight line drawn to include all the points brighter and to the left of the $A_V=0$ mag isochrone.

As visible in the center and right panels of the same figure, most of the contaminants are expected to be located at positive colors and at magnitudes fainter than $m_{130}\sim 15$ mag.
As discussed in more detail in Paper~I, simulations of the expected background extragalactic contaminants and of the expected fore- and background galactic contaminants are used to substantiate the claim that in our selected region there are expected to be very few to no contaminants (Figure~\ref{fig:CMDs}, center). 
The simulations of the extragalactic contaminants are based on CANDELS data \citep{2011ApJS..197...36K}, modeled using spectral energy distribution fitted with the procedure described in \cite{2016ApJ...832...79P}. For the galactic contaminants we used the Besan\c{c}on model of the Galaxy \citep{2003A&A...409..523R}. In both cases we computed the synthetic magnitudes in our WFC3/IR bands using the \texttt{synphot}~\citep{2018ascl.soft11001S} and \texttt{stsynphot}~\citep{lim_pey_lian_2018_3247832} software packages\footnote{Available at \url{https://synphot.readthedocs.io/en/latest/} and \url{http://stsynphot.readthedocs.io/en/latest/}} .

The histograms in the right panel of Figure~\ref{fig:CMDs} show that the simulated contaminant population can account for all the objects outside the fitting region, thus not only is our selection of ONC members pure, but also very complete.
Detailed membership criteria for individual objects are given in Paper~I. 
There we used the Bayes Ratio of the probability of a star being a member divided by the probability of a star belonging to the background.
In the current work we prefer to use a yes/no selection criterion based on whether a simulated star falls within or outside the green-shaded region. We chose this option because of practical implementation considerations.
Effectively, but not exactly, we are using the region where the Bayes Ratio of Paper~I in greater than one.

\subsubsection{Total number of stars}
\label{sec:totnum}

The number of observed stars in the CMD fitting region is itself a realization of a random variable.
The underlying random variable is the mean number of stars, $\Lambda$, that are generated in the ONC within the surveyed footprint.
$\Lambda$ can be regarded as the total intensity of a Poisson Point Process, a class of statistical models used to describe the occurrencies and distributions of points in an observable space. \cite{2015ApJ...808...45G} lay the framework for using Poisson Point Processes in CMD analysis.

In practical terms, if the IMF is normalized to 1 and treated as a probability distribution function, the intensity $\Lambda$ can be seen as the scaling factor in front of IMF and it determines the total number of generated stars, while the details of the IMF shape determine their masses.
Instead of using the full mass range, 0 to infinity, we limit ourselves to the range available in the models: 0.001 to 1.4 M$_\odot$. 

Within our approach the number of stars born in the ONC in that mass interval is a realization of a Poisson distribution with mean equal to $\Lambda$, i.e. $N_{true} \sim Pois(\Lambda)$.
%The number of observed stars within the CMD fitting region is itself a realization of a Poisson Process, which is the \textit{thinned} version of the one which intensity $\Lambda$ \citep[see again][]{2015ApJ...808...45G}. {\bf [qui forse occorre un po' di spiegazione; lambda era un numero di stelle, ora e' una intensita'...]} 
Given the selection function (a combination of incompleteness and the definition of the fit region of ONC members in the CMD), fewer stars are going to make into the CMD, i.e. $N_{obs} \leq N_{true}$.

$\Lambda$ is a fit variable in our MCMC approach. For each MCMC iteration, given the value of $\Lambda$, we extract a $N_{true}$ from a Poisson distribution with mean $\Lambda$. We then iterate all the steps described in Sections \ref{sec:models}-\ref{sec:CMFFR} $N_{true}$ times to obtain an observed CMD.
Given the nature of our likelihood function (see Section~\ref{sec:likelihood}), the number of observed stars in a synthetic CMD, and not $\Lambda$, converges to the number of observed data points, while $\Lambda$ converges to the total number of objects in the 0.001 to 1.4 M$_\odot$ mass range born in the surveyed area, observed ones plus undetected ones.

\subsubsection{Samples Reweighting}
In our fitting approach, we do not actually perform a CMD realization on-the-fly at each MCMC iteration. We rather simulate a very large number of stars, 20 milions, beforehand. This list is used as as a look-up table to draw from, in order generate the CMDs. 
The 3D coordinates of the 20 milions simulated stars are drawn from the spatial distribution described in Section~\ref{sec:spdistr}. Their system masses are drawn from a very broad distribution covering the whole $[0.001, 1.4]$ M$_\odot$ mass interval: a log-normal with $(\mu_S,\sigma_S) = (0.07, 0.7)$, where the footer \textit{S} denotes the sampling distribution.
This broad distribution allows proper sampling of our mass interval edges.
We adopt a $BF_S = 0.2$, or 20\% binary fraction for simulating this list. Thus each of the 50 milion draws has a probability $1-BF_S$ of being single and $BF_S$ of being a binary.
These stars go through the whole process described in Sections~\ref{sec:stpar}--\ref{sec:totnum} to get heir noisy magnitudes (or non-detections) assigned.

In the fitting steps (see Section~\ref{sec:fit}), after drawing the IMF parameters $\theta_{IMF}$, binary fraction $BF$ and intensity value, $\Lambda$, we select stars by reweighting the sample of 50 milion stars according to:
\begin{align}
\mathrm{Single\, stars:} w_i = & \frac{p(m|\theta_{IMF})}{p_{LN}(m|\mu_S,\sigma_S)} * \frac{1-BF}{1-BF_S}\nonumber\\
\mathrm{Binary\, stars:} w_i = & \frac{p(m|\theta_{IMF})}{p_{LN}(m|\mu_S,\sigma_S)} * \frac{BF}{BF_S}
\end{align}
and then choosing the necessary number of stars, $N\sim \mathrm{Pois}(\Lambda)$, from this  list using the $w_i$-s as weights.

The 50 milions stars in the sampling distribution have ages uniformly distributed between 1 and 3 Myr. 
The stellar models have however a coarse spacing of 1 Myr, thus for each extracted age we assign the star to the closest isochrone.
When simulating CMDs with a star formation history with a single peak at 1Myr, we reweight the samples by assigning 0 weight to all stars with age greater than 1.5 Myr, which would be assigned to the 2 or 3 Myr isochrones. Stars with $w_i =0$ are obviously never drawn from the sampling list.

\subsection{Fitting technique}
\label{sec:fit}

We aim at finding the IMF parameters that produce a CMD within the fitting region which minimizes the differences between the observed and simulated data. Our fitting procedure consists in:
\begin{itemize}
    \item defining a functional form of the IMF. This form determines the distribution of the masses of the stars used to populate the CMD. The IMF parametrization choices are discussed in Section~\ref{sec:IMFpar}. Together with the IMF parameters, we fit for the intensity and the binary fraction. In some versions of the fit we also fit for nuisance parameters representing unknown sources of uncertainty (see Section~\ref{sec:NP}).
    \item defining priors on the fit parameters (see Section~\ref{sec:priors}).
    \item defining a likelihood function to compare data and simulations (see Section~\ref{sec:likelihood}).
    \item adopting an algorithm for sampling the posterior distribution of the fit parameters (see Section \ref{sec:MCMC}).
\end{itemize}. 

\subsubsection{IMF parameterization}
\label{sec:IMFpar}
We adopt two different parameterizations of the IMF: a 3-parts, continuous, broken power law, and a log-normal with a high-mass power law tail. These two parametrization allow direct comparison with the descriptions on the Milky Way disk IMF proposed by 
\cite{2001MNRAS.322..231K} and \cite{2003PASP..115..763C}, respectively.
By using these parameterizations, we can assess which form of the disk stellar IMF can better reproduce the high-mass  (M-dwarfs) end of our data, and test whether it can be extrapolated into the brown dwarf and planetary mass regime while still providing a good fit to the data.

The IMF probability distribution functions, are parameterized as follows. For the broken power law:
\begin{align}
p_{\mathrm{BPL}}(m|\alpha_0, \alpha_1,\alpha_2, m_{01}, m_{12})  \propto & \;m^{\alpha_2},\; m\geq m_{12} \nonumber\\
& \; k_{12} m^{\alpha_1},\; m_{01} \leq m < m_{12} \nonumber\\
& \; k_{01} m^{\alpha_0},\; m < m_{01} \nonumber\\
\mathrm{with:} & \nonumber\\& \;  k_{12} = m_{12}^{\alpha_2-\alpha_1}\nonumber\\
& \;  k_{01} = k_{12}m_{01}^{\alpha_1-\alpha_0}\label{eq:bpl}
\end{align}
Here the $\alpha$'s represent the slopes within each segment, and the $m_{ij}$ are the break or transition masses between different segments, or regimes, in solar mass units.  The $k_{ij}$ constants ensure continuity at the break points. The footers run from the low mass end, 0, to the high mass end, 2.
For reference, the values reported by \cite{2001MNRAS.322..231K} are:
\begin{multline}
(\alpha_0, \alpha_1,\alpha_2, m_{01}, m_{12})_{Kroupa} = \\  (-0.3,-1.3,-2.3,0.08,0.5)
\end{multline}

For the log-normal parameterization we have:
\begin{align}
p_{\mathrm{LN}}(m|m_c, \sigma, m_{hm},\alpha_{hm} )  \propto & \; m^{\alpha_{hm}}, m> m_{hm}  \nonumber \\
& \; k\frac{1}{m} e^{-\frac{1}{2}\left(\frac{\log m-\log(m_c)}{\sigma}\right)^2 \label{eq:ln}} \nonumber\\
\mathrm{with:} & \nonumber\\
& \; k = m_{hm}^{\alpha_{hm}+1} e^{\frac{1}{2}\left(\frac{\log m_{hm}-\log(m_c)}{\sigma}\right)^2}.
\end{align}
Here $m_c$ is the characteristic mass in solar mass units (the peak in logarithmic space), $\sigma$ is the width of the log-normal, $m_{hm}$ is the transition mass (in solar units) into the "high mass" regime where the IMF is described as a power-law of slope $\alpha_{hm}$.
For the \cite{2003PASP..115..763C} Galactic Disk IMF we have:
\begin{multline}
(m_c, \sigma, m_{hm},\alpha_{hm} )_{Chabrier} = \\ (0.22,0.57,1,-2.35)
\end{multline}
i.e., the transition to the power law regime is fixed at the Sun's mass, and the high mass slope to the \cite{1955ApJ...121..161S} value.

The normalization constant for each form of the IMF is obtained by numerical integration over the $[0.001, 1.4]$ M$_\odot$ range.

\subsubsection{The priors}
\label{sec:priors}
We impose several priors on the IMF parameters.
For all cases, we impose that all transition masses (from one slope to the other or from the proper log-normal to the high-mass power-law tail) are contained in the [0.001,1.4] M$_\odot$ interval.
For the broken power law case, we impose that the transition masses are sorted, i.e., $m_{01} < m_{12}$. We also impose that the slope must become more positive, i.e. shallower, for progressively lower-mass intervals, i.e. $\alpha_2 \leq \alpha_1 \leq \alpha_0$.
For the log-normal case, we impose that the transition mass to the power law regime is larger than the log-normal peak mass.
We also impose uniform priors on the binary fraction (between 0 and 1), and an improper uniform prior on the total intensity, i.e. that $\Lambda > 0$.

\subsubsection{The likelihood}
\label{sec:likelihood}

In order to compare simulations and observation, we utilize the luminosity function in the F130N band.
We do not adopt a full 2D approach, fitting the observed data distribution on a grid of CMD cells, 
mostly because we do not have enough data to sample the most critical regions of the CMD.

\begin{comment}
Even though they are not truly independent, we treat the two distributions as such, by writing the following likelihood:
\begin{eqnarray}
\label{eq:likelihood}
 \mathcal{L}(obs|\theta) & = & \prod_{j=1}^{N_{bins}(col)} Poisson(n_{obs, col_j} | n_{mod, col_j}(\theta)) \\
 & &  \nonumber\\
 & \times & \prod_{j=1}^{N_{bins}(mag)} Pois(n_{obs, mag_j} | n_{mod, mag_j}(\theta)) \nonumber
\end{eqnarray}
\end{comment}

In our case, the likelihood function is thus defined as:
\begin{equation}
\label{eq:likelihood}
 \mathcal{L}(obs|\theta)  = \prod_{j=1}^{N_{bins}(mag)} Pois(n_{obs, mag_j} | n_{mod, mag_j}(\theta)) 
\end{equation}

where $j$ runs over the bins of the histogram of $m_{130}$. The footers \textit{obs} and \textit{mod} stand for the observed counts, and the synthetic ones, respectively.
The histograms bin sizes are determined using the Freedman Diaconis Estimator, an outlier resistant estimator that takes into account data variability and data size.

In equation (\ref{eq:likelihood}) $\theta$ represents the whole set of fitting parameters, including the IMF parameters, the binary fraction, the total intensity $\Lambda$, and the additional nuisance parameters that account for the excess in color and magnitude spread (see Section~\ref{sec:NP}).

\subsubsection{MCMC approach}
\label{sec:MCMC}
In order to derive the best set of parameters, $\theta$, for a given IMF parametrization (broken power law or log-normal) we utilize \texttt{emcee}, the \cite{2013PASP..125..306F} Python implementation of the \cite{2010CAMCS...5...65G} affine-invariant sampler.

For a proposed set of parameters, $\theta$, we realize a synthetic CMD, as described in Section~\ref{sec:sumCMD}. We then use the likelihood and priors described in Sections~\ref{sec:priors}-\ref{sec:likelihood} to obtain a posterior probability for $\theta$, and proceed by sampling the parameter space.
We ensure, by trial and error, and a posteriori checks, that the individual chains are properly converged and thinned to retain uncorrelated samples.

\section{Results}
\label{sec:res}
\begin{table*}[]
    \footnotesize
    \centering
    \def\arraystretch{1.5}
    \setlength{\tabcolsep}{0.45em}
    \begin{tabular}{ccc|ccccccccc}
    \hline
\multirow{2}{*}{SFH} & Fit         & Nuis.   & \multirow{2}{*}{$\Lambda$} & \multirow{2}{*}{$\alpha_0$} & \multirow{2}{*}{$\alpha_1$} & \multirow{2}{*}{$\alpha_2$} & \multirow{2}{*}{m$_{01}$[M$_{\odot}$]} & \multirow{2}{*}{m$_{12}$[M$_{\odot}$]} & \multirow{2}{*}{BF} & \multirow{2}{*}{$\sigma_{mag}$} & \multirow{2}{*}{$\sigma_{col}$}
\\ 
                        & BF  & Par. & \\
    \hline
    \hline
\multirow{3}{*}{1 Myr} & \multirow{3}{*}{N} &\multirow{3}{*}{N} & $ 2090.1^{+95.4}_{-97.1}$ &  $ -0.58^{+0.06}_{-0.06}$ &  $ -1.24^{+0.26}_{-0.26}$ &  $ -2.06^{+0.29}_{-0.35}$ &  $  0.16^{+0.05}_{-0.04}$ &  $  0.54^{+0.18}_{-0.14}$ &        \\
 & & & \phantom{ 2090.1}$^{+200.3}_{-198.3}$ &  \phantom{ -0.58}$^{+0.13}_{-0.14}$ &  \phantom{ -1.24}$^{+0.55}_{-0.51}$ &  \phantom{ -2.06}$^{+0.53}_{-1.00}$ &  \phantom{  0.16}$^{+0.12}_{-0.12}$ &  \phantom{  0.54}$^{+0.53}_{-0.28}$ &        \\
 & & & \phantom{ 2090.1}$^{+431.7}_{-297.1}$ &  \phantom{ -0.58}$^{+0.49}_{-0.22}$ &  \phantom{ -1.24}$^{+0.70}_{-1.00}$ &  \phantom{ -2.06}$^{+0.67}_{-2.26}$ &  \phantom{  0.16}$^{+0.19}_{-0.15}$ &  \phantom{  0.54}$^{+0.84}_{-0.36}$ &        \\
\hline
\multirow{3}{*}{1-3 Myr} & \multirow{3}{*}{N} &\multirow{3}{*}{N} & $ 2240.1^{+150.2}_{-140.7}$ &  $ -0.65^{+0.07}_{-0.06}$ &  $ -0.85^{+0.12}_{-0.27}$ &  $ -1.85^{+0.38}_{-0.43}$ &  $  0.13^{+0.14}_{-0.08}$ &  $  0.63^{+0.20}_{-0.14}$ &        \\
 & & & \phantom{ 2240.1}$^{+348.1}_{-294.9}$ &  \phantom{ -0.65}$^{+0.20}_{-0.12}$ &  \phantom{ -0.85}$^{+0.19}_{-0.73}$ &  \phantom{ -1.85}$^{+0.65}_{-1.22}$ &  \phantom{  0.13}$^{+0.33}_{-0.12}$ &  \phantom{  0.63}$^{+0.55}_{-0.26}$ &        \\
 & & & \phantom{ 2240.1}$^{+557.7}_{-442.0}$ &  \phantom{ -0.65}$^{+0.60}_{-0.19}$ &  \phantom{ -0.85}$^{+0.26}_{-1.24}$ &  \phantom{ -1.85}$^{+0.85}_{-2.75}$ &  \phantom{  0.13}$^{+0.52}_{-0.13}$ &  \phantom{  0.63}$^{+0.75}_{-0.38}$ &        \\
\hline
\multirow{3}{*}{1 Myr} & \multirow{3}{*}{N} &\multirow{3}{*}{Y} & $ 2068.2^{+99.0}_{-92.5}$ &  $ -0.59^{+0.06}_{-0.06}$ &  $ -1.35^{+0.33}_{-0.35}$ &  $ -2.11^{+0.27}_{-0.37}$ &  $  0.19^{+0.06}_{-0.05}$ &  $  0.59^{+0.24}_{-0.17}$ &        &  $  0.29^{+0.09}_{-0.07}$ &  $  0.06^{+0.03}_{-0.02}$ \\
 & & & \phantom{ 2068.2}$^{+200.6}_{-203.7}$ &  \phantom{ -0.59}$^{+0.13}_{-0.12}$ &  \phantom{ -1.35}$^{+0.64}_{-0.68}$ &  \phantom{ -2.11}$^{+0.51}_{-1.04}$ &  \phantom{  0.19}$^{+0.14}_{-0.14}$ &  \phantom{  0.59}$^{+0.60}_{-0.31}$ &        &  \phantom{  0.29}$^{+0.18}_{-0.13}$ &  \phantom{  0.06}$^{+0.07}_{-0.04}$ \\
 & & & \phantom{ 2068.2}$^{+373.4}_{-297.1}$ &  \phantom{ -0.59}$^{+0.62}_{-0.16}$ &  \phantom{ -1.35}$^{+0.75}_{-1.17}$ &  \phantom{ -2.11}$^{+0.75}_{-1.94}$ &  \phantom{  0.19}$^{+0.21}_{-0.19}$ &  \phantom{  0.59}$^{+0.80}_{-0.38}$ &        &  \phantom{  0.29}$^{+0.21}_{-0.24}$ &  \phantom{  0.06}$^{+0.12}_{-0.06}$ \\
\hline
\multirow{3}{*}{1-3 Myr} & \multirow{3}{*}{N} &\multirow{3}{*}{Y} & $ 2188.0^{+140.0}_{-132.7}$ &  $ -0.66^{+0.08}_{-0.06}$ &  $ -0.86^{+0.13}_{-0.36}$ &  $ -2.15^{+0.50}_{-0.85}$ &  $  0.17^{+0.19}_{-0.13}$ &  $  0.71^{+0.29}_{-0.17}$ &        &  $  0.30^{+0.11}_{-0.08}$ &  $  0.05^{+0.02}_{-0.02}$ \\
 & & & \phantom{ 2188.0}$^{+325.4}_{-260.1}$ &  \phantom{ -0.66}$^{+0.30}_{-0.11}$ &  \phantom{ -0.86}$^{+0.21}_{-0.83}$ &  \phantom{ -2.15}$^{+0.84}_{-2.27}$ &  \phantom{  0.17}$^{+0.37}_{-0.17}$ &  \phantom{  0.71}$^{+0.60}_{-0.32}$ &        &  \phantom{  0.30}$^{+0.18}_{-0.15}$ &  \phantom{  0.05}$^{+0.05}_{-0.04}$ \\
 & & & \phantom{ 2188.0}$^{+506.1}_{-378.8}$ &  \phantom{ -0.66}$^{+0.67}_{-0.16}$ &  \phantom{ -0.86}$^{+0.29}_{-1.47}$ &  \phantom{ -2.15}$^{+1.10}_{-3.83}$ &  \phantom{  0.17}$^{+0.58}_{-0.17}$ &  \phantom{  0.71}$^{+0.68}_{-0.46}$ &        &  \phantom{  0.30}$^{+0.20}_{-0.23}$ &  \phantom{  0.05}$^{+0.13}_{-0.05}$ \\
\hline
\multirow{3}{*}{1 Myr} & \multirow{3}{*}{Y} &\multirow{3}{*}{N} & $ 2053.0^{+91.9}_{-90.2}$ &  $ -0.57^{+0.06}_{-0.06}$ &  $ -1.36^{+0.30}_{-0.28}$ &  $ -2.13^{+0.28}_{-0.35}$ &  $  0.18^{+0.06}_{-0.05}$ &  $  0.60^{+0.22}_{-0.16}$ &  $  0.34^{+0.19}_{-0.14}$ &        &        \\
 & & & \phantom{ 2053.0}$^{+211.4}_{-191.2}$ &  \phantom{ -0.57}$^{+0.13}_{-0.13}$ &  \phantom{ -1.36}$^{+0.64}_{-0.58}$ &  \phantom{ -2.13}$^{+0.52}_{-1.05}$ &  \phantom{  0.18}$^{+0.13}_{-0.13}$ &  \phantom{  0.60}$^{+0.59}_{-0.29}$ &  \phantom{  0.34}$^{+0.50}_{-0.27}$ &        &        \\
 & & & \phantom{ 2053.0}$^{+368.7}_{-320.0}$ &  \phantom{ -0.57}$^{+0.26}_{-0.21}$ &  \phantom{ -1.36}$^{+0.77}_{-0.85}$ &  \phantom{ -2.13}$^{+0.73}_{-2.18}$ &  \phantom{  0.18}$^{+0.24}_{-0.18}$ &  \phantom{  0.60}$^{+0.76}_{-0.40}$ &  \phantom{  0.34}$^{+0.65}_{-0.34}$ &        &        \\
\hline
\multirow{3}{*}{1-3 Myr} & \multirow{3}{*}{Y} &\multirow{3}{*}{N} & $ 2192.3^{+117.7}_{-121.2}$ &  $ -0.66^{+0.07}_{-0.06}$ &  $ -0.94^{+0.17}_{-0.35}$ &  $ -1.93^{+0.31}_{-0.40}$ &  $  0.17^{+0.15}_{-0.11}$ &  $  0.65^{+0.20}_{-0.13}$ &  $  0.55^{+0.23}_{-0.18}$ &        &        \\
 & & & \phantom{ 2192.3}$^{+285.6}_{-243.1}$ &  \phantom{ -0.66}$^{+0.25}_{-0.12}$ &  \phantom{ -0.94}$^{+0.26}_{-0.76}$ &  \phantom{ -1.93}$^{+0.58}_{-1.05}$ &  \phantom{  0.17}$^{+0.31}_{-0.17}$ &  \phantom{  0.65}$^{+0.53}_{-0.27}$ &  \phantom{  0.55}$^{+0.41}_{-0.34}$ &        &        \\
 & & & \phantom{ 2192.3}$^{+519.7}_{-363.7}$ &  \phantom{ -0.66}$^{+0.59}_{-0.17}$ &  \phantom{ -0.94}$^{+0.33}_{-1.14}$ &  \phantom{ -1.93}$^{+0.90}_{-1.99}$ &  \phantom{  0.17}$^{+0.48}_{-0.17}$ &  \phantom{  0.65}$^{+0.71}_{-0.40}$ &  \phantom{  0.55}$^{+0.44}_{-0.52}$ &        &        \\
\hline
\multirow{3}{*}{1 Myr} & \multirow{3}{*}{Y} &\multirow{3}{*}{Y} & $ 2064.6^{+105.1}_{-103.9}$ &  $ -0.58^{+0.06}_{-0.06}$ &  $ -1.42^{+0.36}_{-0.36}$ &  $ -2.16^{+0.31}_{-0.48}$ &  $  0.20^{+0.07}_{-0.06}$ &  $  0.62^{+0.29}_{-0.19}$ &  $  0.29^{+0.20}_{-0.15}$ &  $  0.27^{+0.09}_{-0.06}$ &  $  0.05^{+0.02}_{-0.02}$ \\
 & & & \phantom{ 2064.6}$^{+208.5}_{-213.9}$ &  \phantom{ -0.58}$^{+0.14}_{-0.12}$ &  \phantom{ -1.42}$^{+0.72}_{-0.79}$ &  \phantom{ -2.16}$^{+0.57}_{-1.52}$ &  \phantom{  0.20}$^{+0.16}_{-0.16}$ &  \phantom{  0.62}$^{+0.65}_{-0.35}$ &  \phantom{  0.29}$^{+0.49}_{-0.25}$ &  \phantom{  0.27}$^{+0.19}_{-0.13}$ &  \phantom{  0.05}$^{+0.05}_{-0.04}$ \\
 & & & \phantom{ 2064.6}$^{+313.3}_{-313.5}$ &  \phantom{ -0.58}$^{+0.41}_{-0.19}$ &  \phantom{ -1.42}$^{+0.84}_{-1.25}$ &  \phantom{ -2.16}$^{+0.80}_{-3.59}$ &  \phantom{  0.20}$^{+0.26}_{-0.19}$ &  \phantom{  0.62}$^{+0.77}_{-0.43}$ &  \phantom{  0.29}$^{+0.67}_{-0.29}$ &  \phantom{  0.27}$^{+0.23}_{-0.25}$ &  \phantom{  0.05}$^{+0.08}_{-0.05}$ \\
\hline
\multirow{3}{*}{1-3 Myr} & \multirow{3}{*}{Y} &\multirow{3}{*}{Y} & $ 2182.7^{+126.9}_{-124.1}$ &  $ -0.65^{+0.08}_{-0.06}$ &  $ -0.87^{+0.13}_{-0.38}$ &  $ -2.02^{+0.38}_{-0.53}$ &  $  0.14^{+0.18}_{-0.11}$ &  $  0.67^{+0.21}_{-0.14}$ &  $  0.43^{+0.25}_{-0.18}$ &  $  0.25^{+0.10}_{-0.07}$ &  $  0.05^{+0.02}_{-0.02}$ \\
 & & & \phantom{ 2182.7}$^{+291.2}_{-258.0}$ &  \phantom{ -0.65}$^{+0.30}_{-0.12}$ &  \phantom{ -0.87}$^{+0.22}_{-0.83}$ &  \phantom{ -2.02}$^{+0.69}_{-1.39}$ &  \phantom{  0.14}$^{+0.35}_{-0.14}$ &  \phantom{  0.67}$^{+0.57}_{-0.26}$ &  \phantom{  0.43}$^{+0.50}_{-0.36}$ &  \phantom{  0.25}$^{+0.20}_{-0.15}$ &  \phantom{  0.05}$^{+0.05}_{-0.03}$ \\
 & & & \phantom{ 2182.7}$^{+571.6}_{-401.5}$ &  \phantom{ -0.65}$^{+0.76}_{-0.18}$ &  \phantom{ -0.87}$^{+0.28}_{-1.38}$ &  \phantom{ -2.02}$^{+0.92}_{-2.72}$ &  \phantom{  0.14}$^{+0.61}_{-0.14}$ &  \phantom{  0.67}$^{+0.72}_{-0.39}$ &  \phantom{  0.43}$^{+0.56}_{-0.43}$ &  \phantom{  0.25}$^{+0.24}_{-0.24}$ &  \phantom{  0.05}$^{+0.10}_{-0.05}$ \\
\hline
\hline
\multicolumn{3}{c|}{\cite{2001MNRAS.322..231K} } & & $-0.3^{+0.7}_{-0.7}$ & $-1.3^{+0.5}_{-0.5}$ & $-2.3^{+0.3}_{-0.3}$ & 0.08 & 0.5  &  & & \\
\hline
    \end{tabular}
    \caption{Summary of the results for the broken power law model in the different cases under study. SFH indicates the type of Star Formation History adopted, a single burst at 1 Myr or a uniform distribution between 1 and 3 Myr. The "Fit BF" column indicates whether the binary fraction has been kept fixed at 20\% (N), or fitted for (Y). The "Nuis. Par." column indicates whether nuisance parameters for the extra error in color and magnitude have been introduced and fitted for. 
    For each case there are three rows showing respectively the 68\%, 95\%, 99\% credible intervals limits. The adopted estimator for the distributions averages in the median.
    The \cite{2001MNRAS.322..231K} values are shown for reference, with errors when available.}
    \label{tab:BPLresults}
\end{table*}

We performed our fitting procedure for a set of assumptions.
Other than the two IMF model scenarios, Broken Power Law (BPL) and log-normal (LN), we have varied the star formation history (SFH) between a single burst and an extended case of uniform SFH between 1 and 3 Myr.
Moreover, we either fixed the binary fraction (BF) to 20\% or let it free to vary determining its value in our MCMC procedure.
Finally, for a subset of runs, we have introduced nuisance parameters that account for the extra variance necessary to improve the agreement between the data and the simulations (see Sect.~\ref{sec:NP} below).

In total we have $2$ (IMF models) $\times 2$ (SFH)  $\times 2$ (BF scenarios) $\times 2$ (Nuisance Parameters scenarios) $= 16$ cases. The results are summarized in Table~\ref{tab:BPLresults} for the BPL sub-cases and Table~\ref{tab:LNresults} for the LN ones. The tables report the median values, which we use as the best-fit estimators, as well as the 68\%, 95\% and 99\% credible intervals for the marginal posterior distributions of each fit parameter.
We define these intervals to be equal-tailed, in the sense that the 68\% credible interval contains 68\% of the total probability, with $(100-68)/2 =16\%$ of the remaining probability on either side, with similar definitions for the 95\% and 99\% intervals.
These credible intervals are analogous to "1-2-3$\sigma$" intervals for a Gaussian distribution, and in the following we might adopt the $n-\sigma$ intervals with some abuse of notation. However, the  posterior distributions of our parameters are generally not Gaussian and not even symmetric in most cases, thus we also explicitly report the interval limits for the fit parameters in Tables~\ref{tab:BPLresults} and \ref{tab:LNresults}. The marginal distributions for the single parameters are useful for defining the parameter credible intervals, but do not capture the whole information available in the full posterior distribution. Strong correlations exist in many cases, e.g. between slope values and transition mass values in the BPL model, or between the $\mu$ and $\sigma$ parameters in the LN one. We capture this information in the corner plots\footnote{Plots produced using the \texttt{corner.py} routine, \cite{corner}} of Figures~\ref{fig:res_BPL_1}--\ref{fig:res_LN_1_bin_n}.

\begin{table*}[]
    \footnotesize
    \centering 
    \setlength{\tabcolsep}{0.45em}
    \def\arraystretch{1.5}
    \begin{tabular}{ccc|ccccccccc}
    \hline
\multirow{2}{*}{SFH} & Fit         & Nuis.   & \multirow{2}{*}{$\Lambda$} & \multirow{2}{*}{$m_c$} & \multirow{2}{*}{$\sigma$} & \multirow{2}{*}{$m_{hm}$} & \multirow{2}{*}{$\alpha_{hm}$} & \multirow{2}{*}{BF} & \multirow{2}{*}{$\sigma_{mag}$} & \multirow{2}{*}{$\sigma_{col}$}
\\ 
                        & BF  & Par. & \\
    \hline
    \hline
\multirow{3}{*}{1 Myr} & \multirow{3}{*}{N} &\multirow{3}{*}{N} & $ 2056.6^{+109.8}_{-109.4}$ &  $  0.26^{+0.11}_{-0.07}$ &  $  1.00^{+0.17}_{-0.12}$ &  $  0.86^{+0.22}_{-0.27}$ &  $ -2.22^{+0.35}_{-0.34}$ &        \\
 & & & \phantom{ 2056.6}$^{+259.2}_{-248.0}$ &  \phantom{  0.26}$^{+0.26}_{-0.10}$ &  \phantom{  1.00}$^{+0.39}_{-0.23}$ &  \phantom{  0.86}$^{+0.47}_{-0.47}$ &  \phantom{ -2.22}$^{+0.80}_{-0.91}$ &        \\
 & & & \phantom{ 2056.6}$^{+477.9}_{-346.8}$ &  \phantom{  0.26}$^{+0.49}_{-0.13}$ &  \phantom{  1.00}$^{+0.82}_{-0.31}$ &  \phantom{  0.86}$^{+0.53}_{-0.59}$ &  \phantom{ -2.22}$^{+1.48}_{-2.20}$ &        \\
\hline
\multirow{3}{*}{1-3 Myr} & \multirow{3}{*}{N} &\multirow{3}{*}{N} & $ 2210.6^{+184.3}_{-118.4}$ &  $  0.47^{+0.16}_{-0.10}$ &  $  1.18^{+0.19}_{-0.12}$ &  $  1.04^{+0.18}_{-0.15}$ &  $ -1.99^{+0.54}_{-0.43}$ &        \\
 & & & \phantom{ 2210.6}$^{+585.5}_{-262.5}$ &  \phantom{  0.47}$^{+0.44}_{-0.20}$ &  \phantom{  1.18}$^{+0.47}_{-0.23}$ &  \phantom{  1.04}$^{+0.31}_{-0.35}$ &  \phantom{ -1.99}$^{+1.17}_{-1.11}$ &        \\
 & & & \phantom{ 2210.6}$^{+1003.5}_{-396.8}$ &  \phantom{  0.47}$^{+0.79}_{-0.25}$ &  \phantom{  1.18}$^{+0.82}_{-0.33}$ &  \phantom{  1.04}$^{+0.36}_{-0.56}$ &  \phantom{ -1.99}$^{+1.72}_{-2.18}$ &        \\
\hline
\multirow{3}{*}{1 Myr} & \multirow{3}{*}{N} &\multirow{3}{*}{Y} & $ 2038.2^{+104.4}_{-98.2}$ &  $  0.31^{+0.13}_{-0.10}$ &  $  1.07^{+0.15}_{-0.14}$ &  $  0.75^{+0.28}_{-0.17}$ &  $ -2.35^{+0.33}_{-0.41}$ &        &  $  0.30^{+0.10}_{-0.08}$ &  $  0.05^{+0.03}_{-0.02}$ \\
 & & & \phantom{ 2038.2}$^{+231.6}_{-217.8}$ &  \phantom{  0.31}$^{+0.27}_{-0.15}$ &  \phantom{  1.07}$^{+0.33}_{-0.26}$ &  \phantom{  0.75}$^{+0.54}_{-0.29}$ &  \phantom{ -2.35}$^{+0.73}_{-1.04}$ &        &  \phantom{  0.30}$^{+0.18}_{-0.16}$ &  \phantom{  0.05}$^{+0.07}_{-0.04}$ \\
 & & & \phantom{ 2038.2}$^{+471.7}_{-321.3}$ &  \phantom{  0.31}$^{+0.40}_{-0.18}$ &  \phantom{  1.07}$^{+0.60}_{-0.33}$ &  \phantom{  0.75}$^{+0.63}_{-0.44}$ &  \phantom{ -2.35}$^{+1.18}_{-2.12}$ &        &  \phantom{  0.30}$^{+0.20}_{-0.28}$ &  \phantom{  0.05}$^{+0.13}_{-0.05}$ \\
\hline
\multirow{3}{*}{1-3 Myr} & \multirow{3}{*}{N} &\multirow{3}{*}{Y} & $ 2195.9^{+175.6}_{-142.8}$ &  $  0.45^{+0.15}_{-0.10}$ &  $  1.19^{+0.14}_{-0.12}$ &  $  1.06^{+0.17}_{-0.18}$ &  $ -2.09^{+0.63}_{-0.73}$ &        &  $  0.32^{+0.10}_{-0.09}$ &  $  0.05^{+0.03}_{-0.02}$ \\
 & & & \phantom{ 2195.9}$^{+484.1}_{-279.7}$ &  \phantom{  0.45}$^{+0.38}_{-0.18}$ &  \phantom{  1.19}$^{+0.33}_{-0.23}$ &  \phantom{  1.06}$^{+0.30}_{-0.39}$ &  \phantom{ -2.09}$^{+1.18}_{-1.95}$ &        &  \phantom{  0.32}$^{+0.17}_{-0.19}$ &  \phantom{  0.05}$^{+0.07}_{-0.04}$ \\
 & & & \phantom{ 2195.9}$^{+943.8}_{-443.6}$ &  \phantom{  0.45}$^{+0.73}_{-0.24}$ &  \phantom{  1.19}$^{+0.77}_{-0.33}$ &  \phantom{  1.06}$^{+0.33}_{-0.62}$ &  \phantom{ -2.09}$^{+1.74}_{-3.85}$ &        &  \phantom{  0.32}$^{+0.18}_{-0.26}$ &  \phantom{  0.05}$^{+0.13}_{-0.05}$ \\
\hline
\multirow{3}{*}{1 Myr} & \multirow{3}{*}{Y} &\multirow{3}{*}{N} & $ 2036.3^{+116.7}_{-110.6}$ &  $  0.24^{+0.11}_{-0.05}$ &  $  0.97^{+0.17}_{-0.12}$ &  $  0.90^{+0.18}_{-0.25}$ &  $ -2.27^{+0.29}_{-0.31}$ &  $  0.31^{+0.18}_{-0.15}$ &        &        \\
 & & & \phantom{ 2036.3}$^{+263.3}_{-233.2}$ &  \phantom{  0.24}$^{+0.27}_{-0.09}$ &  \phantom{  0.97}$^{+0.39}_{-0.21}$ &  \phantom{  0.90}$^{+0.38}_{-0.45}$ &  \phantom{ -2.27}$^{+0.75}_{-0.85}$ &  \phantom{  0.31}$^{+0.43}_{-0.27}$ &        &        \\
 & & & \phantom{ 2036.3}$^{+869.6}_{-370.1}$ &  \phantom{  0.24}$^{+0.46}_{-0.12}$ &  \phantom{  0.97}$^{+0.84}_{-0.30}$ &  \phantom{  0.90}$^{+0.49}_{-0.54}$ &  \phantom{ -2.27}$^{+1.91}_{-1.24}$ &  \phantom{  0.31}$^{+0.64}_{-0.31}$ &        &        \\
\hline
\multirow{3}{*}{1-3 Myr} & \multirow{3}{*}{Y} &\multirow{3}{*}{N} & $ 2087.9^{+110.5}_{-96.0}$ &  $  0.43^{+0.13}_{-0.09}$ &  $  1.15^{+0.15}_{-0.13}$ &  $  0.99^{+0.16}_{-0.17}$ &  $ -2.09^{+0.35}_{-0.34}$ &  $  0.65^{+0.19}_{-0.19}$ &        &        \\
 & & & \phantom{ 2087.9}$^{+258.7}_{-208.6}$ &  \phantom{  0.43}$^{+0.38}_{-0.17}$ &  \phantom{  1.15}$^{+0.40}_{-0.23}$ &  \phantom{  0.99}$^{+0.35}_{-0.38}$ &  \phantom{ -2.09}$^{+0.75}_{-0.88}$ &  \phantom{  0.65}$^{+0.31}_{-0.41}$ &        &        \\
 & & & \phantom{ 2087.9}$^{+602.9}_{-329.4}$ &  \phantom{  0.43}$^{+0.70}_{-0.23}$ &  \phantom{  1.15}$^{+0.75}_{-0.31}$ &  \phantom{  0.99}$^{+0.41}_{-0.55}$ &  \phantom{ -2.09}$^{+1.23}_{-1.66}$ &  \phantom{  0.65}$^{+0.35}_{-0.59}$ &        &        \\
\hline
\multirow{3}{*}{1 Myr} & \multirow{3}{*}{Y} &\multirow{3}{*}{Y} & $ 2028.4^{+108.4}_{-107.2}$ &  $  0.28^{+0.12}_{-0.08}$ &  $  1.04^{+0.15}_{-0.13}$ &  $  0.76^{+0.24}_{-0.20}$ &  $ -2.30^{+0.31}_{-0.32}$ &  $  0.25^{+0.19}_{-0.15}$ &  $  0.29^{+0.10}_{-0.07}$ &  $  0.05^{+0.02}_{-0.02}$ \\
 & & & \phantom{ 2028.4}$^{+238.9}_{-202.2}$ &  \phantom{  0.28}$^{+0.23}_{-0.12}$ &  \phantom{  1.04}$^{+0.33}_{-0.24}$ &  \phantom{  0.76}$^{+0.48}_{-0.35}$ &  \phantom{ -2.30}$^{+0.69}_{-0.94}$ &  \phantom{  0.25}$^{+0.45}_{-0.23}$ &  \phantom{  0.29}$^{+0.18}_{-0.15}$ &  \phantom{  0.05}$^{+0.06}_{-0.04}$ \\
 & & & \phantom{ 2028.4}$^{+475.8}_{-303.3}$ &  \phantom{  0.28}$^{+0.42}_{-0.15}$ &  \phantom{  1.04}$^{+0.53}_{-0.33}$ &  \phantom{  0.76}$^{+0.63}_{-0.43}$ &  \phantom{ -2.30}$^{+1.22}_{-2.01}$ &  \phantom{  0.25}$^{+0.69}_{-0.24}$ &  \phantom{  0.29}$^{+0.20}_{-0.26}$ &  \phantom{  0.05}$^{+0.11}_{-0.05}$ \\
\hline
\multirow{3}{*}{1-3 Myr} & \multirow{3}{*}{Y} &\multirow{3}{*}{Y} & $ 2083.7^{+109.2}_{-106.1}$ &  $  0.40^{+0.12}_{-0.08}$ &  $  1.12^{+0.14}_{-0.11}$ &  $  1.02^{+0.16}_{-0.16}$ &  $ -2.27^{+0.44}_{-0.45}$ &  $  0.60^{+0.22}_{-0.22}$ &  $  0.27^{+0.10}_{-0.08}$ &  $  0.05^{+0.03}_{-0.02}$ \\
 & & & \phantom{ 2083.7}$^{+314.0}_{-207.6}$ &  \phantom{  0.40}$^{+0.33}_{-0.16}$ &  \phantom{  1.12}$^{+0.34}_{-0.21}$ &  \phantom{  1.02}$^{+0.32}_{-0.34}$ &  \phantom{ -2.27}$^{+0.97}_{-1.17}$ &  \phantom{  0.60}$^{+0.36}_{-0.45}$ &  \phantom{  0.27}$^{+0.19}_{-0.17}$ &  \phantom{  0.05}$^{+0.07}_{-0.04}$ \\
 & & & \phantom{ 2083.7}$^{+671.1}_{-302.1}$ &  \phantom{  0.40}$^{+0.63}_{-0.24}$ &  \phantom{  1.12}$^{+0.69}_{-0.30}$ &  \phantom{  1.02}$^{+0.38}_{-0.52}$ &  \phantom{ -2.27}$^{+1.72}_{-2.06}$ &  \phantom{  0.60}$^{+0.40}_{-0.58}$ &  \phantom{  0.27}$^{+0.22}_{-0.24}$ &  \phantom{  0.05}$^{+0.13}_{-0.05}$ \\
\hline

\multicolumn{3}{c|}{\cite{2003PASP..115..763C} (Singles)} & & $0.079^{-0.016}_{+0.021}$ & $0.69^{-0.01}_{+0.05}$ & $1.0$ & $-2.3$&  &  & &     \\
\multicolumn{3}{c|}{\cite{2003PASP..115..763C} (Systems)} & & $0.22$ & $0.57$ & $1.0$ & $-2.3$&  &  & &     \\
\hline
    \end{tabular}
    \caption{Summary of the results for the log-normal model in the different cases under study. Columns meanings are discussed in Table~\ref{tab:BPLresults}. The \cite{2003PASP..115..763C} values are shown for reference, with errors when available.}
    \label{tab:LNresults}
\end{table*}

\subsection{Overall assessment of the results quality}
\begin{figure*}[t]
\begin{center}
\includegraphics[width=0.825\textwidth]{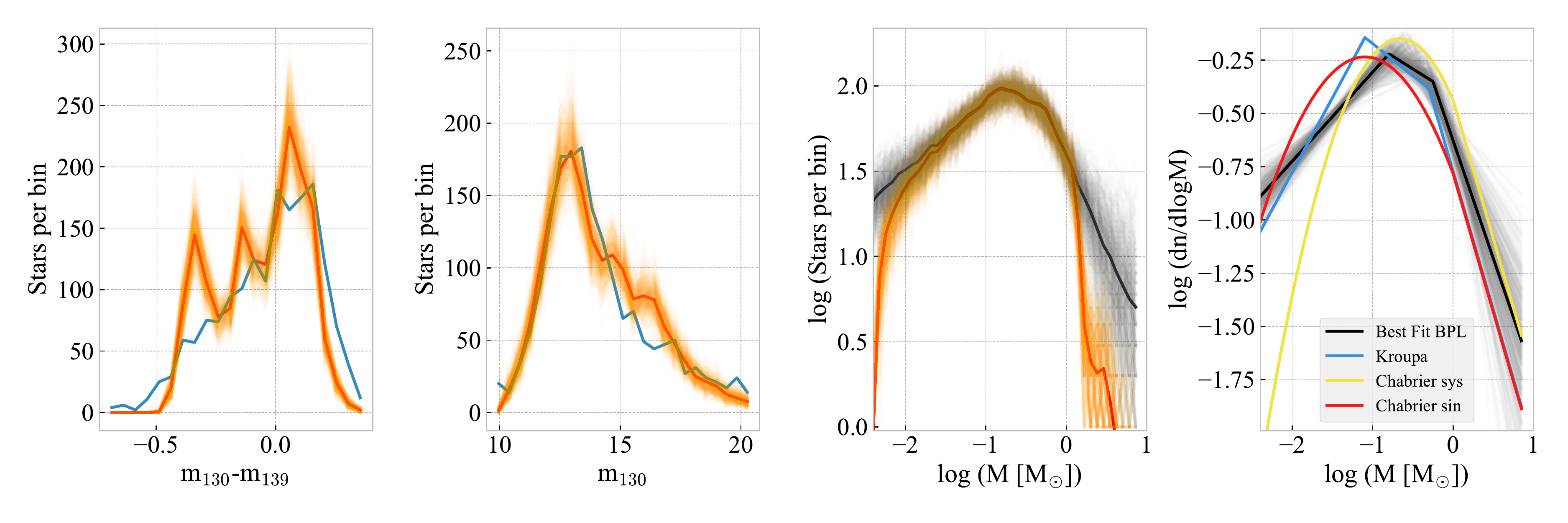}
\includegraphics[width=0.825\textwidth]{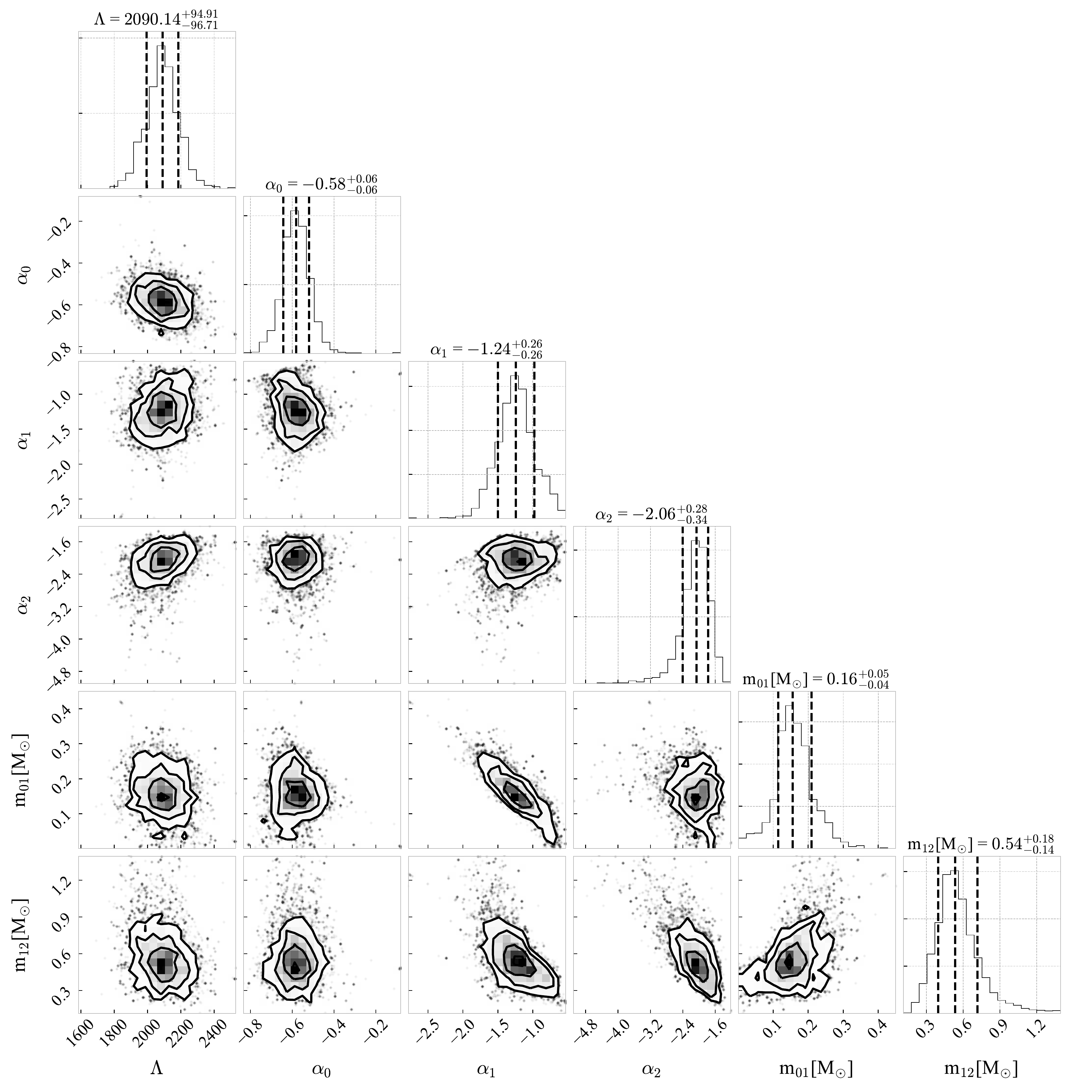}
\caption{Broken Power Law IMF fit results, 1Myr case. Top row, left and center-left: color and magnitude distributions of the observed stars (cyan), with superimposed the color and magnitude distributions from the best fit IMF parameters (red), and from 1000 draws from the posteriors (yellow-orange). Note the overall good fit to the luminosity function. The sharp peaks in the simulated color distribution function show however that the simulations underestimate the existing color spread.
Top row, center-right panel: masses drawn from the best-fit IMF (black) and from 1000 samples from the posterior (gray). Red/yellow: corresponding observable masses within the CMD fitting region, note the high-mass drop due to our high-mass saturation cutoff, and the low-mass drop due to incompleteness. Top row, right panel, comparison of the best-fit IMF with the standard \cite{2001MNRAS.322..231K} and \cite{2003PASP..115..763C} Galactic Disk IMFs.
Bottom: corner plot for the fit parameters. The median and the limits of the 68\% credible interval ($1\sigma$) are reported for each variable.\label{fig:res_BPL_1}
} 
\end{center}
\end{figure*}

\begin{figure*}[t]
\begin{center}
\includegraphics[width=0.825\textwidth]{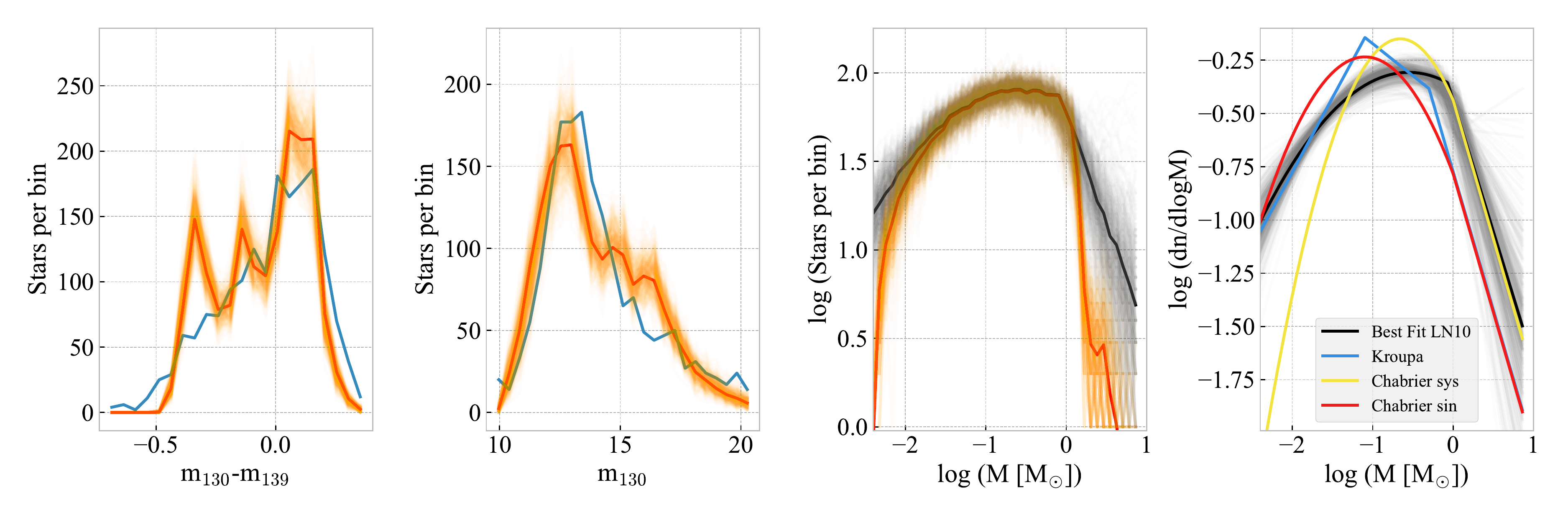}
\includegraphics[width=0.825\textwidth]{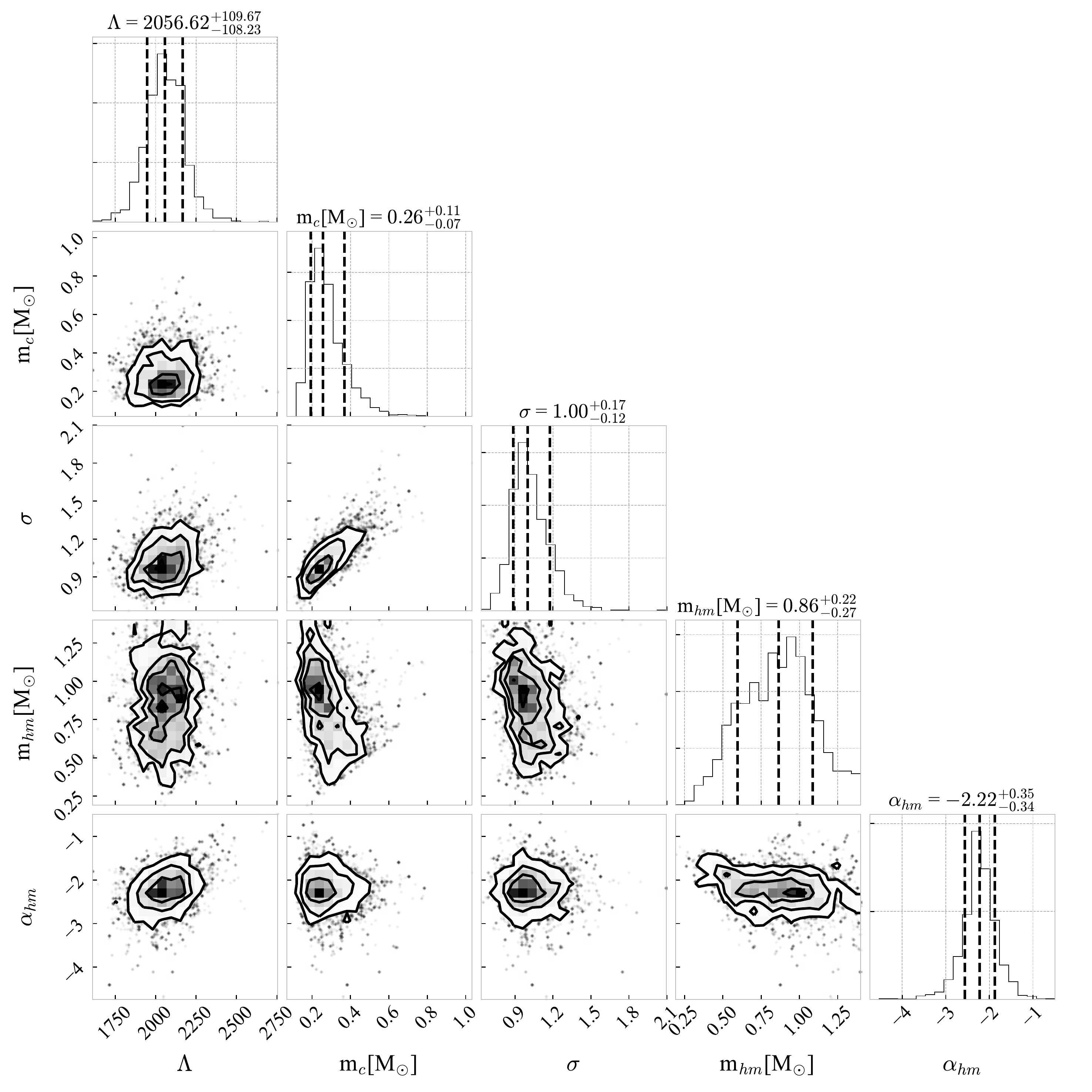}
\caption{log-normal IMF fit results, 1Myr burst SFH case . See Figure~\ref{fig:res_BPL_1} for an explanation.\label{fig:res_LN_1}
} 
\end{center}
\end{figure*}

For a first assessment of the level of success of our procedure, we can consider the simple case of the BPL model, with an assumed 1~Myr single-burst SFH and with a fixed BF of 20\%.
The corresponding fit results are shown in Figure~\ref{fig:res_BPL_1}.
Overall the simulated luminosity function reproduces the observed one very well (center-left panel). There is however some disagreement in the left panel, where the simulated $m_{130}-m_{F139}$ color distribution shows three sharp peaks that are not visible in the observations. These peaks correspond to the vertical regions of the model isochrone, around colors $m_{130}-m_{F139} \simeq -0.4,-0.1$ and $0.05$~mag. 
It is worth remembering that while we are simulating CMDs, we are fitting only for the luminosity function. Therefore, the color distribution has a minor impact on the results of the fit, it just provides us with a sanity check. In the case of this simple model, the check is not satisfactory. 
Still, for the case under consideration, the fit parameters values are overall in good agreement with the \cite{2001MNRAS.322..231K} IMF, specially in the high (indicated by a "2") and intermediate (indicated by a "1") mass ranges, while differences are present in the lowest mass range (indicated by a "0"):
the very low-mass slope is slightly steeper $\alpha_0=-0.58$, but still well within 1$\sigma$ of their $\alpha_0=-0.3 \pm 0.7$. 
Moreover, the mid-to-low transition mass is found to be $m_{01} = 0.16$ M$_\odot$, compared to 0.08 M$_\odot$ of \cite{2001MNRAS.322..231K}. This difference, for the given assumptions, is marginally significant, with 0.08 M$_\odot$ outside the 68\% credible interval but within the 95\% one.

A similar level of agreement between the data and the simulated luminosity function is found for the log-normal case, again for a 1~Myr SFH and with the BF held fixed at 20\%, as visible in Figure~\ref{fig:res_LN_1}. As in the BPL case, the simulations fail at reproducing the color distribution.
The characteristic mass, $m_c = 0.26$~M$_\odot$ is very similar to the value reported by \cite{2003PASP..115..763C} for their system mass function, 0.22~M$_\odot$, while
the LN width $\sigma = 1.00$ is larger by a significant amount than their 0.57 value. This value falls outside the 99\% credible interval for our estimate.

Our wider distribution is needed to predict the number of very low mass object observed in the ONC. This discrepancy in the LN case echoes our finding that in the BPL case the low-mass power law declines more slowly than the \cite{2001MNRAS.322..231K} value. In practice, for both IMF models the galactic disk IMF canonical parameters seem to under-predict the number of very low mass objects in the young ONC system.

\subsection{Broadening the simulated CMDs: the star formation history}
\begin{figure*}[t]
\begin{center}
\includegraphics[width=0.825\textwidth]{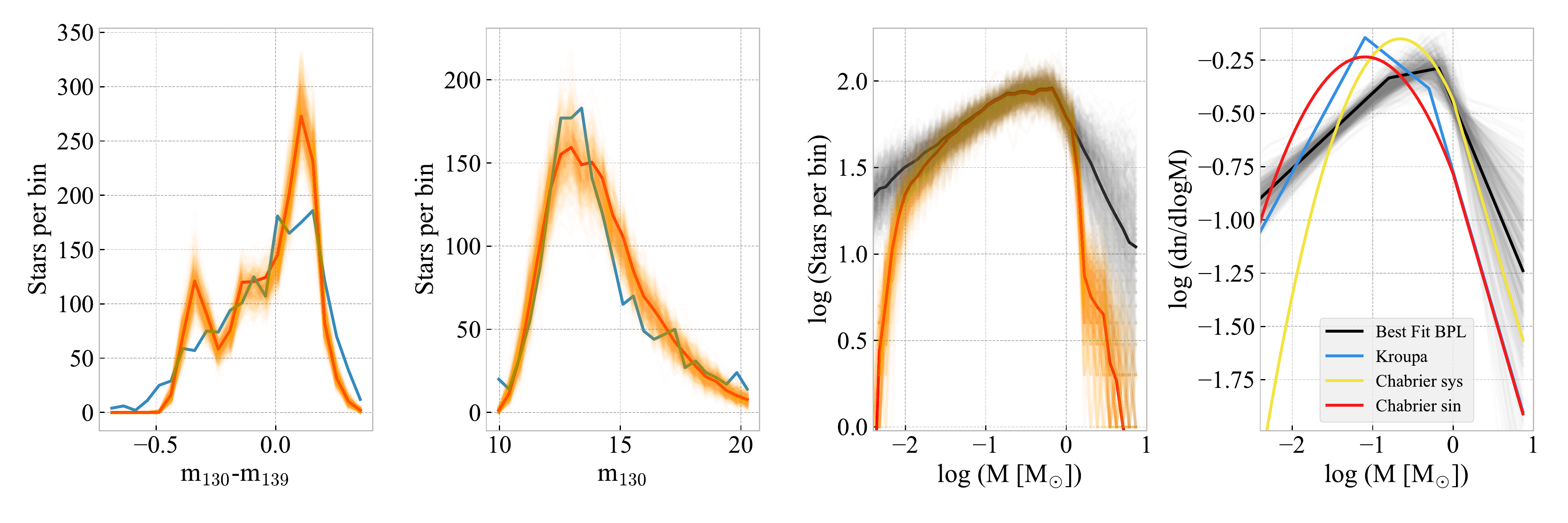}
\includegraphics[width=0.825\textwidth]{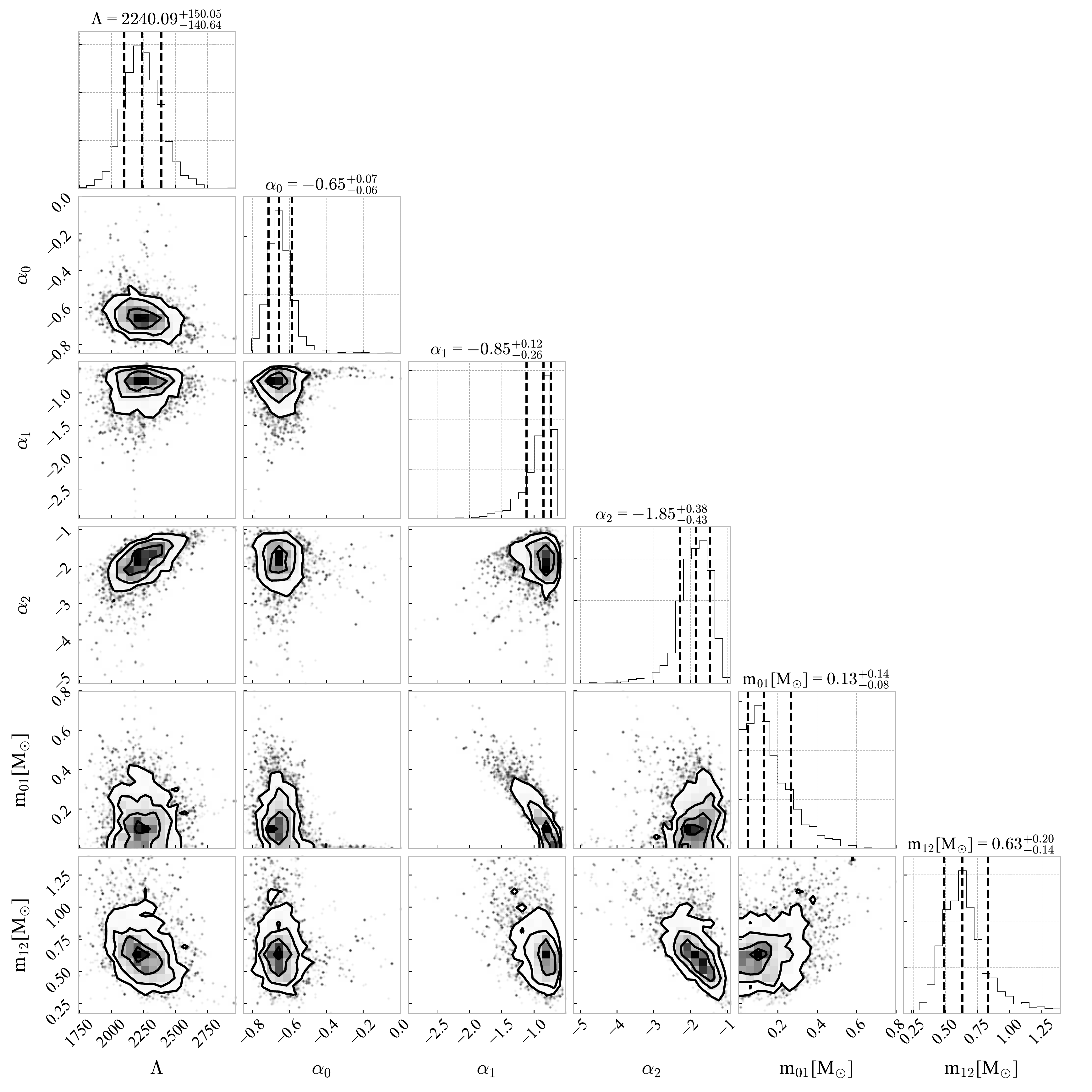}
\caption{BPL IMF fit results, 1-to-3 Myr uniform SFH case. See Figure~\ref{fig:res_BPL_1} for an explanation.\label{fig:res_BPL_1to3}
} 
\end{center}
\end{figure*}

 Before claiming a significant difference in the predicted counts at very low mass with respect to the widely adopted parameterizations of the Galactic Disk IMF, a better agreement between the observed and the simulated color distributions is desirable. The  discrepancy warrants further investigation. To this purpose, we analyze what happens when a broader SFH is adopted. It turns out that a uniform SFH between 1 and 3 Myr partly smooths out the color distribution of simulated stars with respect to the single burst case. This is shown for the BPL model in Figure~\ref{fig:res_BPL_1to3} (the LN model, not shown, gives similar results). While improved, the agreement in the color distribution is still not entirely acceptable.
 
Changing the SFH generally impacts the derived parameter values, as detailed in Tables~\ref{tab:BPLresults} and \ref{tab:LNresults}. It is worth noticing that for every sub-case, the 1 Myr burst always gives results that are more aligned with the galactic disk IMF results with respect to the 1-3 Myr uniform case.
For example, in the BPL model case the slope in the middle mass range, $\alpha_1$, is quite different from the \cite{2001MNRAS.322..231K} value of $-1.3$ when we adopt a 1-to-3 Myr uniform SFH, with best fit values in the $-0.85$ to $-0.95$ range, while we obtain $\alpha_1$ of $-1.2$ to $-1.4$ for a 1~Myr SFH.
Similarly, the characteristic mass in LN case is shifted from $0.2$--$0.3$~M$_\odot$ of the 1 Myr SFH case, to $0.4$--$0.5$~M$_\odot$ for the 1-to-3 Myr case, further from the \cite{2003PASP..115..763C} system IMF value of $0.22$~M$_\odot$.

Our uniform SFH assumption can be regarded as an extreme attempt at reconciling the observed and simulated color distributions. 
Several papers indicate that the ONC region may have undergone extended star formation within the past 1-3 Myr. For example, \cite{2011A&A...534A..83R} find that part of the observed luminosity spread in ONC pre-main sequence stars may be explained by star formation activity between 1.5 and 3.5 Myr ago. \cite{2017A&A...604A..22B} reported evidence of three multiple pre-main sequence populations in the ONC, implying a possible "bursty" SFH at 1, 2, and 3 Myr, with the 3 Myr population being the most prominent. This result is corroborated by the analysis of \cite{2019A&A...627A..57J} who rule out binaries as the explanation for the multiple sequences observed by \cite{2017A&A...604A..22B} and confirm the latter paper trimodal age distribution.
Our adopted uniform SFH across 3 Myr, although maximizing the predicted spread in luminosity even with respect to the bursty solution by \cite{2017A&A...604A..22B}, still cannot explain the observed difference between models and data. It must be noted that while we observe the well known luminosity spread, we do not find signatures of multiple sequences in our infrared data, thus a uniform distribution seems appropriate for modeling them.
Nevertheless, given that it does not help fully reconciling our model with the data, and given that it makes the main derived parameters deviate from the galactic IMF even in the well studied 0.2 -- 0.5~M$_\odot$ regime, we shall adopt the 1~Myr model over the 1-to-3 Myr one as our benchmark in the rest of this paper. Nevertheless, we report the results for both scenarios.

\subsection{Broadening the simulated CMDs: the binary fraction}
\begin{figure*}[t]
\begin{center}
\includegraphics[width=0.825\textwidth]{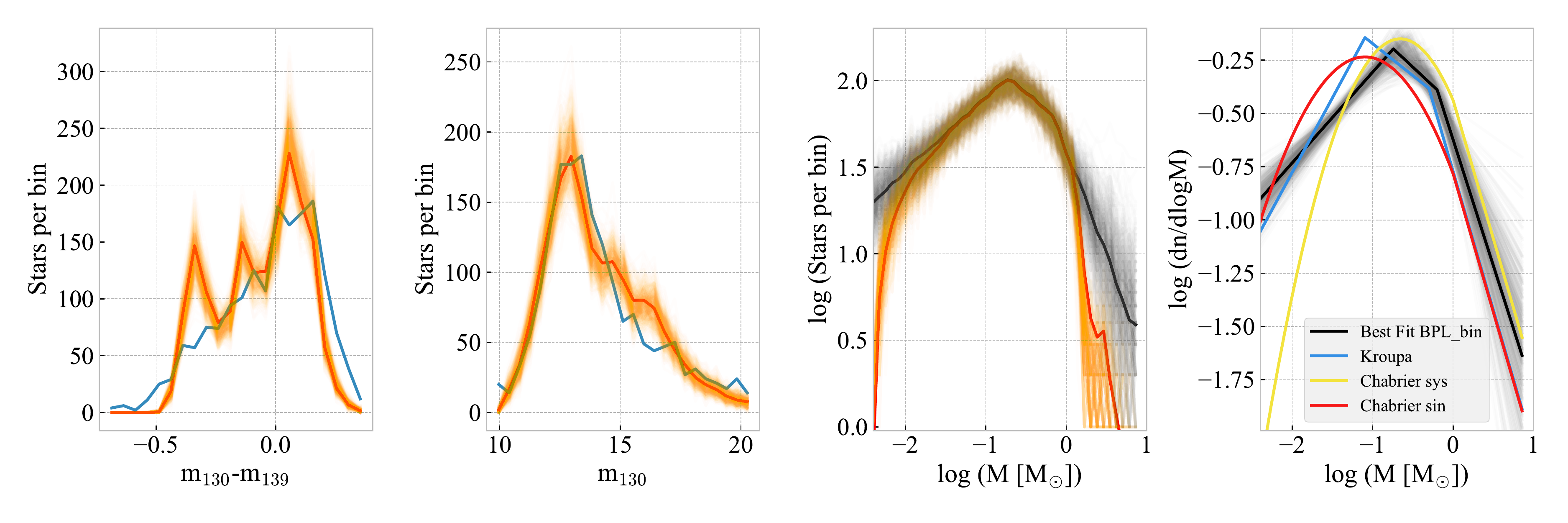}
\includegraphics[width=0.825\textwidth]{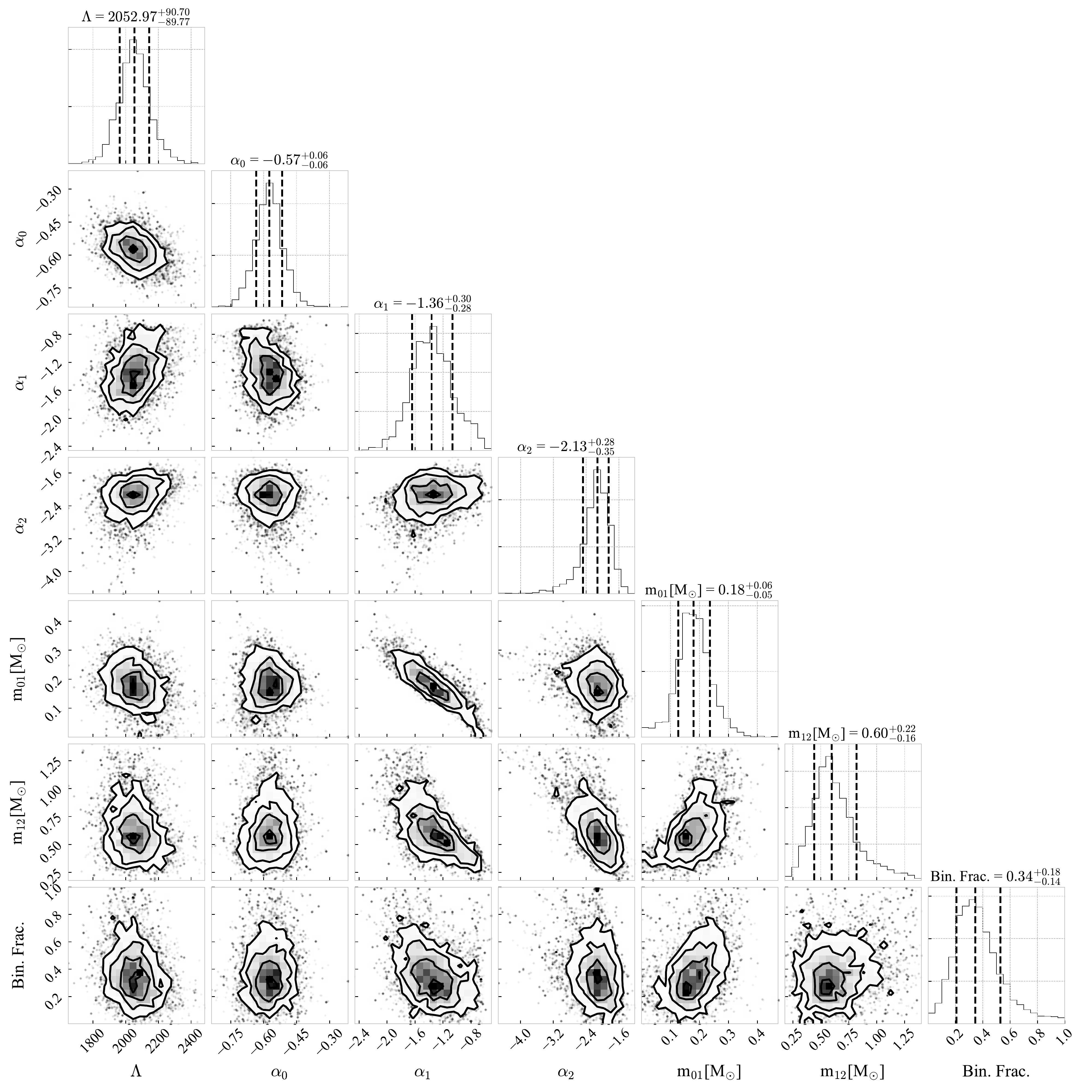}
\caption{BPL IMF fit results, 1 Myr burst SFH case, with binary fraction left as a free fit parameter. See Figure~\ref{fig:res_BPL_1} for an explanation.\label{fig:res_BPL_1_bin}
} 
\end{center}
\end{figure*}

A further possibility to spread out the stars in color in the CMD is to allow the binary fraction to vary.
There is no strong a priori reason for forcing the binary fraction to 20\%, except for convenience and a possible similarity with respect to the galactic disk, where the fraction is observed to be around this number \citep{1991A&A...248..485D,2010ApJS..190....1R}.
However, even leaving the binary fraction free does not solve the color distribution problem, as visible in Fig.~\ref{fig:res_BPL_1_bin} for the BPL model and the 1~Myr SFH case (the LN model, not shown, gives similar results). This is due first to the fact that the best fit binary fraction value, 0.34 is not very different from 0.2, the latter value being within the 68\% credible interval (for the analogous LN case the best fit binary fraction is 0.31). Moreover, specially at low masses, the isochrones are quite vertical thus the color difference between primary and secondary in a binary is small, contributing little to smoothing the color distribution.
Even though it does not solve the color distribution problem, we favor model scenarios where the binary fraction is a free parameter. As visible in Tables~\ref{tab:BPLresults} and \ref{tab:LNresults}, including the binary fraction in the fit has only minor impact on the other parameters.

\subsection{Broadening the simulated CMDs: nuisance parameters}
\label{sec:NP}
\begin{figure*}[t]
\begin{center}
\includegraphics[width=0.925\textwidth]{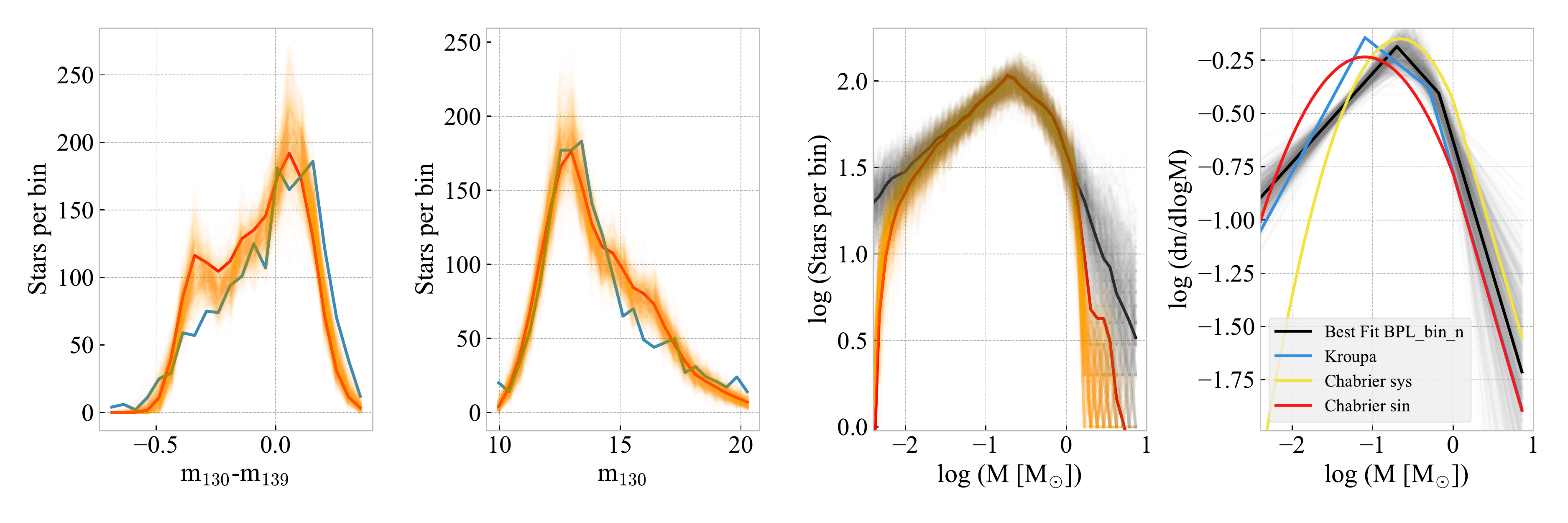}
\includegraphics[width=0.925\textwidth]{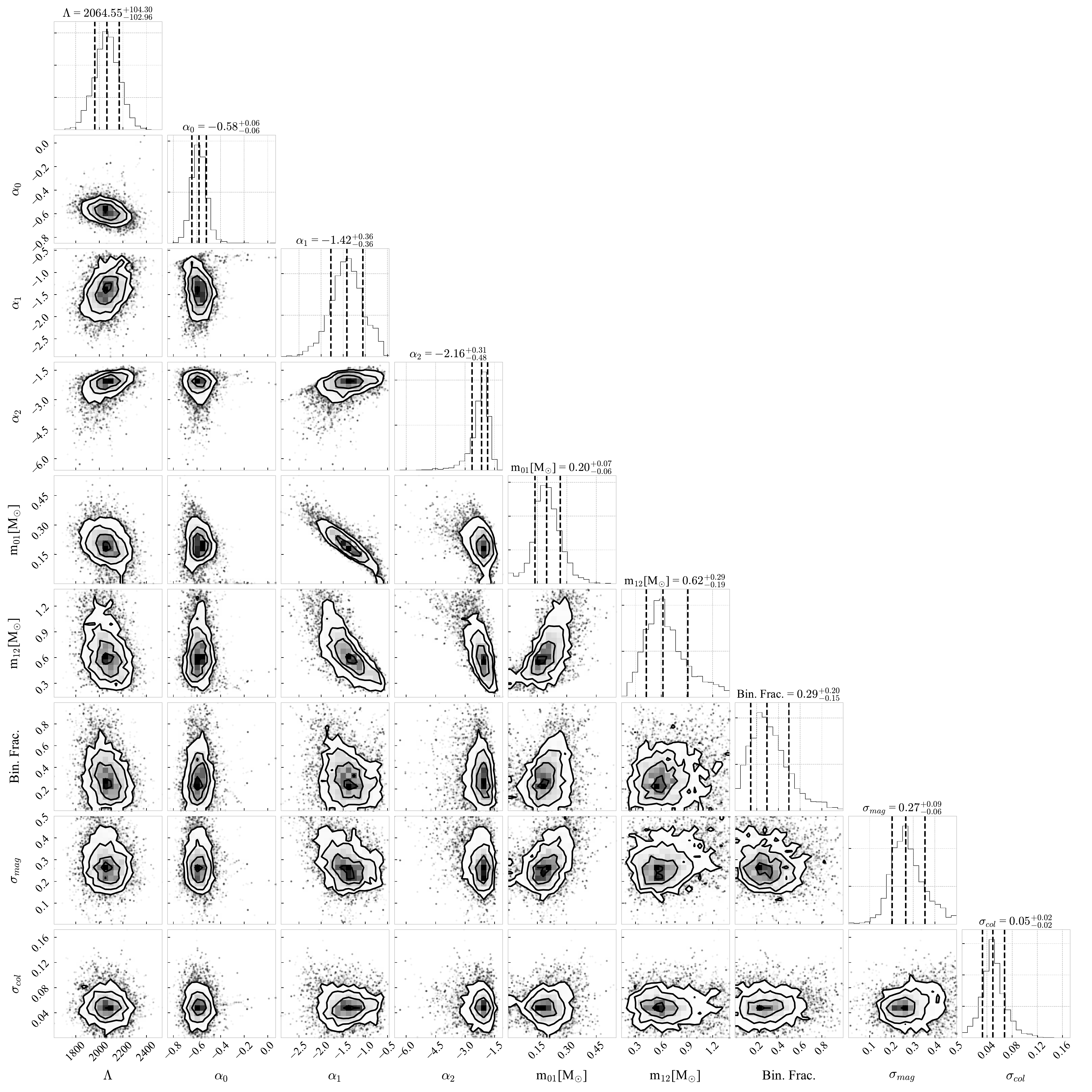}
\caption{BPL IMF fit results, 1 Myr burst SFH case, with binary fraction left as a free fit parameter and including nuisance paramters. See Figure~\ref{fig:res_BPL_1} for an explanation\label{fig:res_BPL_1_bin_n}} 
\end{center}
\end{figure*}

The fact that neither binaries, nor the IMF functional form or the assumed SFH help smooth the simulated color distribution, implies that some ingredients are missing in the adopted stellar models.
We know, in fact, than young pre-main sequence stars in clusters show broad luminosity and temperature ranges, that are often interpreted as age-spreads, but could in fact be caused by a series of phenomena such as accretion, presence of disk, magnetic activity, among others \citep[see e.g.][]{2011MNRAS.418.1948J,2011A&A...534A..83R,2015ApJ...807..174S}.
Our stellar models include only the light from the stellar photosphere and thus cannot predict the luminosity and colors variations associated to these phenomena.

\begin{figure*}[t]
\begin{center}
\includegraphics[width=0.925\textwidth]{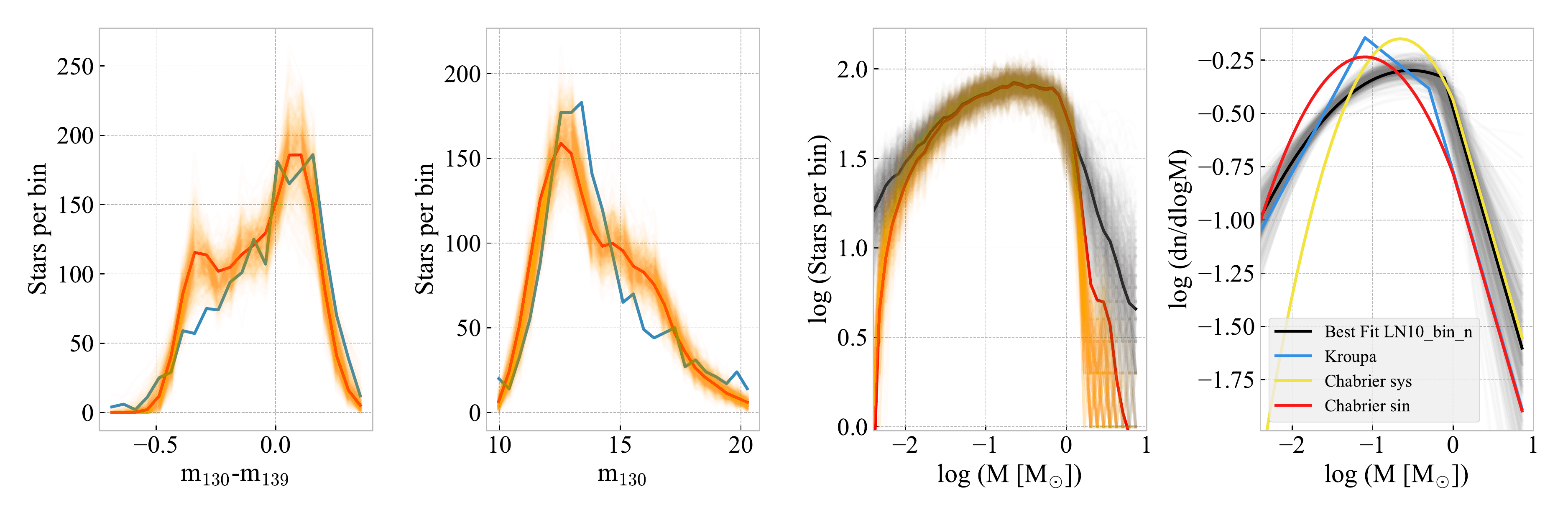}
\includegraphics[width=0.925\textwidth]{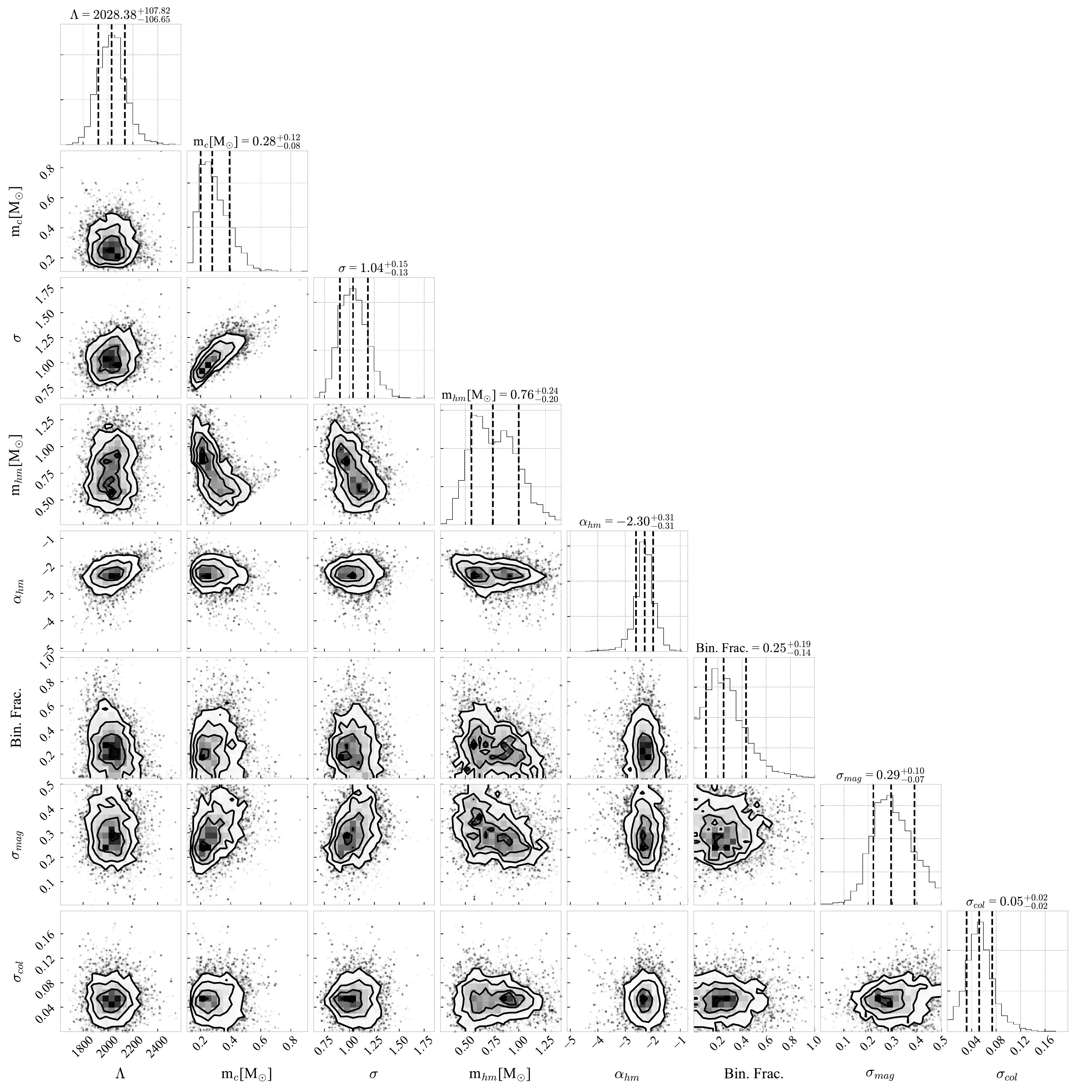}
\caption{LN IMF fit results, 1 Myr burst SFH case, with binary fraction left as a free fit parameter. See Figure~\ref{fig:res_BPL_1} for an explanation\label{fig:res_LN_1_bin_n}} 
\end{center}
\end{figure*}

To circumvent this problem, we introduce two additional parameters: two unknown additional photometric scatters, one in magnitude, one in color, parameterized as the $\sigma$-s of two Gaussian distributions. We fit for these nuisance parameters simultaneously with the IMF parameters, the binary fraction, and the total number of stars. Introducing the nuisance parameters allows us to obtain a color and luminosity distribution that resembles the observed ones to a much better degree with respect to all the other cases mentioned above.
The results are shown in Figures~\ref{fig:res_BPL_1_bin_n} and \ref{fig:res_LN_1_bin_n} for the BPL and LN case respectively. In both cases we show only the 1Myr SFH sub-case. For the reasons mentioned in Section~\ref{sec:discuss}, we favor this scenario to the uniform 1-to-3 Myr SFH.

At this stage we do not speculate on the true origin of the additional spread captured by our nuisance parameters.
We note that the IMF fitting results are never profoundly affected by the addition of the nuisance parameters, thus our results on the IMF are not strongly dependent on them. 
This is an obvious consequence of the fact that we are only using the luminosity function in our likelihood calculation, and that the luminosity function was already well reproduced without nuisance parameters.
At the same time, the color distribution agreement when the nuisance parameters are introduced gives us more confidence in the final results.

%\begin{figure*}[t]
%\begin{center}
%\includegraphics[width=0.925\textwidth]{Compare_results_BPL_bin_n_1to3Myr_onlylum}
%\includegraphics[width=0.925\textwidth]{CornerPlots_BPL_bin_n_1to3Myr_onlylum}
%\caption{Same as Figure~\ref{fig:resbasic} but for a Broken Power Law IMF, 1 to 3 Myr star %formation history and with nuisance parameters\label{fig:res_BPL_1to3_bin_n}} 
%\end{center}
%\end{figure*}

%\begin{figure*}[t]
%\begin{center}
%\includegraphics[width=0.925\textwidth]{Compare_results_LN10_bin_n_1to3Myr_onlylum}
%\includegraphics[width=0.925\textwidth]{CornerPlots_LN10_bin_n_1to3Myr_onlylum}
%\caption{Same as Figure~\ref{fig:resbasic} but for a log-normal IMF, 1 to 3 Myr star %formation history and with nuisance parameters\label{fig:res_LN_1to3_bin_n}} 
%\end{center}
%\end{figure*}

\subsection{A sanity check}
Although our analysis is designed to approach the IMF fitting problem in the most rigorous terms, probing the multiple sources of uncertainty and deriving a robust estimate of the IMF parameters posterior probability distribution function, it is appropriate to assess our findings using a simplified, more conventional approach, with the caveats illustrated below.

\begin{figure*}[t]
\begin{center}
\includegraphics[width=0.925\textwidth]{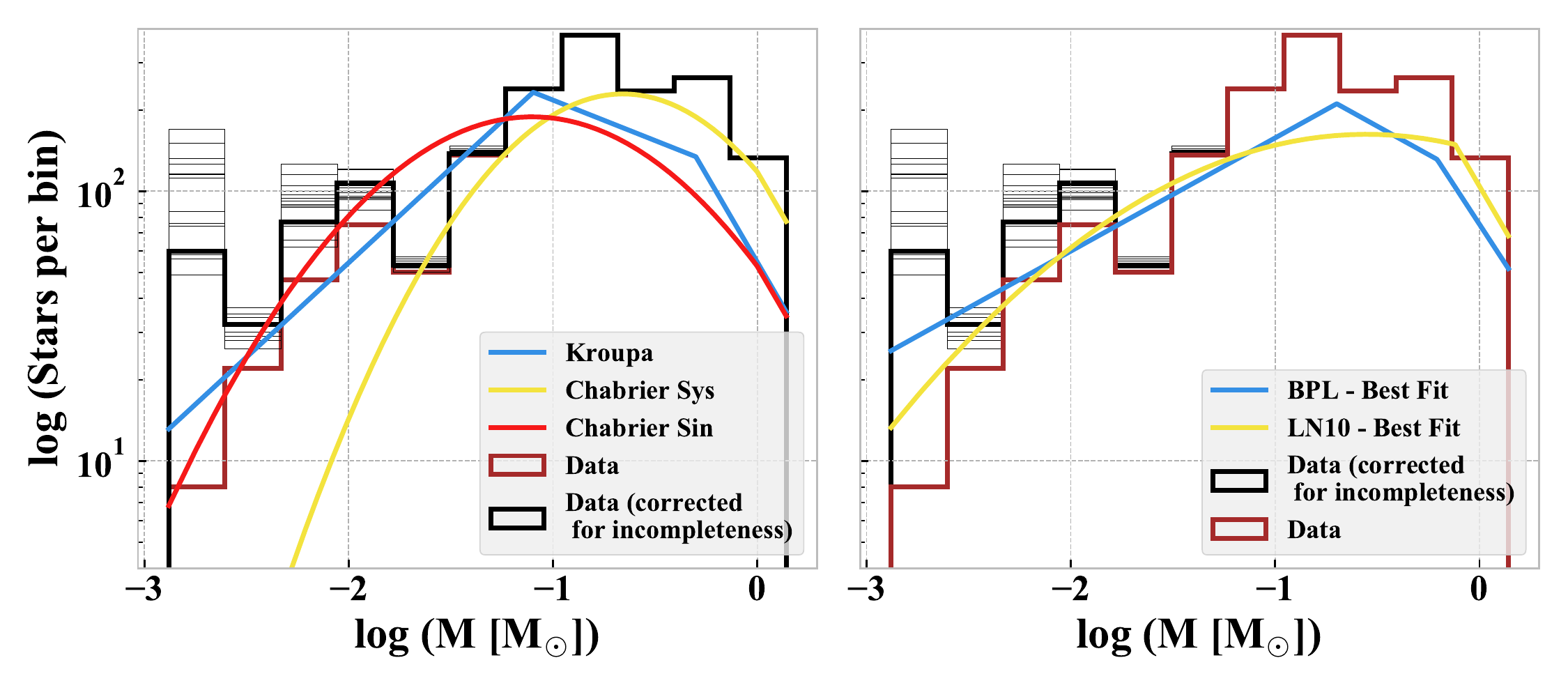}
\caption{Histograms of the individual masses in the fit region using the catalog from Paper~I. The data are shown in brown. For the completeness corrected data, we show in black the correction corresponding to using the average number of detection failures from the corresponding negative binomials (see text for details), while the lighter histograms correspond to 15 actual random draws from the negative binomial, to give an idea of the correction uncertainty. Left and right panels are separated for clarity showing the IMF Disk canonical models on the left and our best fit on the right, using a single burst 1Myr SFH. \label{fig:basicIMF}} 
\end{center}
\end{figure*}

We begin by adopting the mass estimates for the individual sources in our fit region from Paper~I. We remind the reader that those masses are derived by translating the individual stars in the CMD along the reddening vector up to the zero-reddening 1 Myr isochrone, i.e. "dereddening' the measured magnitudes, and reading out the mass-intercept along the isochrone.
These masses are shown in the histogram of Fig.~\ref{fig:basicIMF}.

Together with the histogram of observed masses, we show the same histogram corrected for completeness, applied as follows:
\begin{itemize}
\item we consider the measured magnitudes for each star in the histogram $(m_{130}, m_{139})_i$ and its position $(RA, Dec)_i$
\item we use the completeness maps from Paper I, which give us an average completeness in concentric rings from the ONC center (see their Figure~6)
\item given $(RA, Dec)_i$ we determine in which concentric ring the star was found and we adopt the corresponding completeness curves
\item we interpolate the completeness curves for F130N and F139M at the measured $(m_{130}, m_{139})_i$
\item given that detection in each band is treated independently in our catalog, we obtain a single completeness value by multiplying the interpolated values for each band. We thus have a number $0<\mathcal{C}_i<1$ for star $i$
\item we adopt a negative binomial distribution $NB(n=1,\,p={C}_i)$. A negative binomial of parameters $(n,p)$ represents the probability distribution of the number of "failures" necessary to obtain a number $n$ of successes if $p$ is the probability of a single success. Here $n=1$ represents the success of having measured one time our star $i$, and $p={C}_i$  -the completeness- represents the probability of that single succes.
\item we extract a number of failures for each observed star. This number corresponds to the extra stars added to the completeness-corrected histogram of Fig.~\ref{fig:basicIMF}
\end{itemize}

Fig.~\ref{fig:basicIMF}, left panel shows our data before and after completeness correction, together with the canonical Galactic Disk IMFs by \cite{2001MNRAS.322..231K} and \cite{2003PASP..115..763C}. We show our best fit results separately in the right panel of the Figure, to avoid confusion (the data histograms are the same). %The figure shows the average correction as well as individual draws for the negative binomial failure rates. 
The canonical models tend to underpredict the decline at low masses and also underestimate the location of the peak (Chabrier-systems IMF) or the power-law mass break (Kroupa IMF), consistently with what  demonstrated the previous sections by using  the rigorous forward modeling approach.
Our best fit values (shown for the case of a single episode of star formation at 1 Myr) are more in line with the data, confirming the reliability our procedure.

We take the occasion to remark a sometimes overlooked detail in the widely used approach of correcting a histogram of e.g. measured magnitudes, or inferred masses, for completeness.
Completeness is the probability of a star of given intrinsic magnitude $\mathcal{M}_\mathrm{in}$ to be observed, not the number of failed attempts necessary before measuring a star of given $\mathcal{M}_\mathrm{out}$. Given the noise model in the observation in question, the measured magnitude will be $\mathcal{M}_\mathrm{out} = \mathcal{M}_\mathrm{in}+ \epsilon$. Even in the case of symmetric noise (e.g. $\epsilon \sim N(0,\sigma)$, gaussian noise) this relation cannot be inverted. Again, completeness is not the number of missed stars at $\mathcal{M}_\mathrm{out}$. This may be true in the region where completeness is absolutely perfect, i.e. 100\% detection probability, but it is not true in general. In the region where completeness monotonically decreases, which is typically the region where completeness corrections comes into play, the fact that there is a noise term, $\epsilon$, makes a difference.  There is no way of knowing what the $\mathcal{M}_\mathrm{in}$ corresponding to a measured $\mathcal{M}_\mathrm{out}$ is, and thus the completeness fraction as measured, e.g. using artificial stars experiments, cannot be applied to the output magnitude, unless one is implicitly adopting $\mathcal{M}_\mathrm{out}$ = $\mathcal{M}_\mathrm{in}$. This difference is subtle, but substantial in regions of rapidly declining completeness, where usually $\epsilon$ starts to be significant. 

Although in this subsection we followed this common approach for the sake of illustration, we warn against trying to "enhance" or "complete" observed histograms. On the contrary, one should more safely "de-complete", for a lack of a better word, the predicted counts from models by using intrinsic magnitudes predicted by the models, knowing the probability of a star of a given magnitude of being observable, as well as the actual magnitude (statistically) at which that star will be observed.

\section{Discussion}
\label{sec:discuss}

%As shown in Tables~\ref{tab:BPLresults} and \ref{tab:BPLresults}, the best fit parameters depend slightly on the assumed star formation history. 
%A pure 1~Myr old population requires a 24\% binary fraction, lower than the 42\% derived using a uniform age distribution between 1 and 3 Myr. For comparison \cite{2010ApJS..190....1R} estimate that $33\%\pm2\%$ of the solar-type stars within 25~pc from the Sun have at least a stellar or substellar companion. In the case of the Orion Nebula cluster, thius value drops to $8.8\%\pm 1.1\%$ limited to the distance range 67.5 to 675 AU \citep{2007AJ....134.2272R}. 
%We do not over interpret the numbers on the binary fraction and just comment that rather than fixing the binary fraction to some value, we prefer letting it vary, and thus account for its unknown value an let it propagate into the other variables of the problem.
%Changing the star formation history also affects the derived IMF parameters. All the variables of the problem, binary fraction included, are strongly correlated, thus this outcome is not surprising. {\bf [questo punto della star formation history non sembra discusso...]}

As motivated in Sect.~\ref{sec:res}, our favored scenarios are those that adopt a 1~Myr burst of star formation, leave the binary fraction free to vary and utilize nuisance parameters to better reproduce the observed color distribution.
The results for these scenarios are shown in Figures~\ref{fig:res_BPL_1_bin_n} and \ref{fig:res_LN_1_bin_n}.
We note that both the BPL and the LN IMF forms allow us to reproduce the observed color and luminosity distribution very well. 

For the BPL case, our low-mass slope, $\alpha_0 = -0.58$, is shallower than  \cite{2001MNRAS.322..231K}, $-0.3$, a value which is outside our 95\% credible interval. However the \cite{2001MNRAS.322..231K} $\alpha_0$ has a very large uncertainty of $\pm 0.7$.
We also predict a higher transition mass between the mid and low ranges, $m_{01} = 0.20$~M$\odot$, with respect to $0.08$~M$\odot$ of \cite{2001MNRAS.322..231K}. For $m_{01}$ the inconsistency is only marginal, with the \cite{2001MNRAS.322..231K} value outside our 68\% credible interval but within the 95\% one.
The depth of our data allows constraining our parameter values in the very low mass range to a much higher precision than \cite{2001MNRAS.322..231K}, as demonstrated by our much smaller uncertainties (our 99\% credible interval for $\alpha_0$ is narrower than their $1\sigma$ interval). It is thus not surprising that with such superior dataset we have slightly different results, but within the uncertainties of the older ones. 

For the LN case the agreement of the fit parameters with the work of \cite{2003PASP..115..763C} is slightly worse.
This reflects what already discussed in Sect.~\ref{sec:res} for the basic scenario without nuisance parameters and with the binary fraction fixed at 20\%.
The right case to compare with is the \cite{2003PASP..115..763C} system mass function, which accounts for the presence of unresolved binaries, which reflect our ONC case. The \cite{2003PASP..115..763C} system mass function and our LN fit have similar peak masses, our value being 0.28~M$_\odot$ and theirs, 0.22~M$_\odot$, being within our 68\% credible interval. At the same time the best fit LN that we derive has a much larger $\sigma$ (it is broader) to account for the number of low mass objects that is otherwise underpredicted by the \cite{2003PASP..115..763C} system mass function.
The \cite{2003PASP..115..763C} $\sigma$ value is outside our 99\% credible interval for $\sigma$.

Our discussion on the impact of unknown model uncertainties and different parameterization of the SFH, should however warn the reader that the overall accuracy of our results could be affected by systematic errors in these assumptions.
We also remind the reader that the stellar models we adopted become less reliable in the planetary mass regime, and require a color correction (see Paper~I) to reproduce the observed water absorption strength. Therefore we might expect a change in some of the fit details, if different models are adopted.
It is similarly true that any attempt to measure the IMF must deal with the conversion of magnitudes to masses using stellar models, and with assumptions on the SFH. The latter, for the Galactic Disk case, is more complex and thus more prone to systematic errors than that of a young cluster like the ONC. While highlighting our own systematic problems we thus remind the reader that those problems exist in all other IMF derivations.

\begin{comment}
In particular the \cite{2001MNRAS.322..231K} IMF predicts slightly more low-mass objects at the edge of our mass range, than the \citep{2003PASP..115..763C} system IMF (see the rightmost panels at the top of Figures~\ref{fig:res_BPL_1_bin_n}--\ref{fig:res_LN_1_bin_n}). To compensate for this, our log-normal fits tend to have a larger $\sigma$, i.e. a broader low-mass tail than the \citep{2003PASP..115..763C} results. In other words, the observed number of planetary mass objects is well predicted by a \cite{2001MNRAS.322..231K} IMF, but slightly underpredicted by a \cite{2003PASP..115..763C} one.
Yet, given that most of these differences come from the very low mass end of our sample, where incompleteness treatment may start to play a role in the exact parameters derivation, we do not claim that these differences are significant {\bf [non possiamo piutttosto elencare i caveats e basta?]}.
\end{comment}

In conclusion, we have that both \cite{2001MNRAS.322..231K} and \citep{2003PASP..115..763C} models underpredict the number of very low mass objects with respect to our best fit models. However the description by \cite{2001MNRAS.322..231K} is closer and more consistent with our results.
Yet, in terms of goodness of fit, the BPL and LN models do an equally good job in reproducing the observed luminosity function, i.e. our best fit models in either parametrization are equivalent and are good fits to the data.

We observe that the \textit{stellar} IMF extends continuously into the brown dwarfs and planetary mass objects regime. This could imply that a common mechanism is required to explain the formation of such objects. Rather than being the result of an entirely different formation scenario, low mass objects in the field -even below to the Deuterium burning limit- constitute the tail of the stellar mass distribution, and can be thus regarded like ordinary stars from the initial formation mechanism standpoint.

We underline that no secondary peak of the IMF is observed at low masses as claimed for example by \cite{2002ApJ...573..366M} and more recently by \cite{2016MNRAS.461.1734D}.
%that a secondary peak of the IMF must exist at 0.025 M$_\odot$, with the primary peak at the \cite{2003PASP..115..763C} or \cite{2010AJ....139.2679B} location of $\sim 0.25$~M$_\odot$.
%We do not find this peak in our data, and e
%Even without introducing the nuisance parameters that improve the accuracy of our color fit by adding extra degrees of freedom,
Our data show that the observed luminosity function is fully compatible with a simple single-peaked log-normal or a qualitatively similar 3 parts broken power law.
Using ground-based near-infrared broad-band data to infer the shape of the low-mass IMF in region as complex as the ONC is prone to major uncertainties related to the contamination by background sources. In particular, the irregular column density of the Orion Molecular Cloud allows to detected faint, spurious sources that cannot be reliably accounted for using statistical methods.% We speculate that the \cite{2016ApJ...818...59D} results may have been affected by an incomplete subtraction of background objects.% Our 
%are superior in that they allow to build a truly pure sample of ONC members without the need of statistical subtraction, but from the water absorption induced blue color in the $F130N-F139M$ \textit{HST} filters directly.  

\section{Summary and Conclusions}
\label{sec:sumandconc}

We use \textit{HST} WFC3/IR to build a pure sample of Orion Nebula Clusters members.
Using the  data from  Paper~I and synthetic CMD modeling we derive the best fit IMF parameters for a broken power law and log-normal model.

Both of these parameterizations allow us to recover the observed luminosity function well.
Our best fit parameters are in qualitative agreement with the \cite{2001MNRAS.322..231K} ones for the broken power law model and with the \cite{2003PASP..115..763C} ones for the log-normal, specially above the stellar/sub-stellar mass boundary.
Nevertheless, both these canonical models tend to underpredict the counts in the very low mass regime.
There is however a slight preference for the \cite{2001MNRAS.322..231K} models, while a broader log-normal than \cite{2003PASP..115..763C} is needed to fit the low-end of our observation. 

Our results do not confirm the existence of a secondary very-low mass peak of the IMF as reported by \cite{2016MNRAS.461.1734D}.
On the contrary, we show that an extension to lower masses of the Galactic Disk IMF provides a very good description of the data, i.e. the canonical description of the distribution of masses extends well into the brown dwarfs and planetary mass regime.

We interpret the continuous, gradual variation of the IMF from the stellar to the sub-stellar regime down to few Jupiter masses
as the manifestation of a single formation scenario, regulated by a common set of physical processes acting through the full mass spectrum.
%of stellar and mass spectrum including free-floating planetary mass objects.

By allowing to probe the main molecular bands in the near- and mid-IR photospheric spectra, space-based observations have unique power disentangling  brown dwarfs and planetary-mass objects from the main galactic population. The analysis carried out in this paper therefore illustrates the potential of future JWST observations, both in imaging and spectroscopy, in reconstructing the IMF to Jupiter masses in galactic clusters of different age, size and metallicity.

\acknowledgments
Support for Program number GO-13826 was provided by NASA through a grant from the Space Telescope Science Institute, which is operated by the Association of Universities for Research in Astronomy, Incorporated, under NASA contract NASS-26555. The authors are very grateful to and would like to thank Prof. Lynne Hillendbrand (Caltech) and Dr. Laurent Pueyo (STScI) for very useful discussion along the preparation of this manuscript and Dr. David Soderblom for carefully reading through it.

\bibliography{biblio_OrionIMF}

\begin{thebibliography}{}
\expandafter\ifx\csname natexlab\endcsname\relax\def\natexlab#1{#1}\fi
\providecommand{\url}[1]{\href{#1}{#1}}

\bibitem[{{Allard} {et~al.}(2012){Allard}, {Homeier}, \&
  {Freytag}}]{2012RSPTA.370.2765A}
{Allard}, F., {Homeier}, D., \& {Freytag}, B. 2012, Philosophical Transactions
  of the Royal Society of London Series A, 370, 2765

\bibitem[{{Allison} \& {Goodwin}(2011)}]{2011MNRAS.415.1967A}
{Allison}, R.~J., \& {Goodwin}, S.~P. 2011, \mnras, 415, 1967

\bibitem[{{Asplund} {et~al.}(2009){Asplund}, {Grevesse}, {Sauval}, \&
  {Scott}}]{2009ARA&A..47..481A}
{Asplund}, M., {Grevesse}, N., {Sauval}, A.~J., \& {Scott}, P. 2009, \araa, 47,
  481

\bibitem[{{Baraffe} {et~al.}(2015){Baraffe}, {Homeier}, {Allard}, \&
  {Chabrier}}]{2015A&A...577A..42B}
{Baraffe}, I., {Homeier}, D., {Allard}, F., \& {Chabrier}, G. 2015, \aap, 577,
  A42

\bibitem[{{Bate}(2009)}]{2009MNRAS.392.1363B}
{Bate}, M.~R. 2009, \mnras, 392, 1363

\bibitem[{{Bate} \& {Bonnell}(2005)}]{2005MNRAS.356.1201B}
{Bate}, M.~R., \& {Bonnell}, I.~A. 2005, \mnras, 356, 1201

\bibitem[{{Beccari} {et~al.}(2017){Beccari}, {Petr-Gotzens}, {Boffin},
  {Romaniello}, {Fedele}, {Carraro}, {De Marchi}, {de Wit}, {Drew}, \&
  {Kalari}}]{2017A&A...604A..22B}
{Beccari}, G., {Petr-Gotzens}, M.~G., {Boffin}, H.~M.~J., {et~al.} 2017, \aap,
  604, A22

\bibitem[{{Bochanski} {et~al.}(2010){Bochanski}, {Hawley}, {Covey}, {West},
  {Reid}, {Golimowski}, \& {Ivezi{\'c}}}]{2010AJ....139.2679B}
{Bochanski}, J.~J., {Hawley}, S.~L., {Covey}, K.~R., {et~al.} 2010, \aj, 139,
  2679

\bibitem[{{Bonnell} {et~al.}(2008){Bonnell}, {Clark}, \&
  {Bate}}]{2008MNRAS.389.1556B}
{Bonnell}, I.~A., {Clark}, P., \& {Bate}, M.~R. 2008, \mnras, 389, 1556

\bibitem[{{Bonnell} {et~al.}(2007){Bonnell}, {Larson}, \&
  {Zinnecker}}]{2007prpl.conf..149B}
{Bonnell}, I.~A., {Larson}, R.~B., \& {Zinnecker}, H. 2007, Protostars and
  Planets V, 149

\bibitem[{{Brown} {et~al.}(2018){Brown}, {Vallenari}, {Prusti}, {de Bruijne},
  {Babusiaux}, {Bailer-Jones}, {Biermann}, {Evans}, {Eyer}, \&
  et~al.}]{2018A&A...616A...1G}
{Brown}, A.~G.~A., {Vallenari}, A., {Prusti}, T., {et~al.} 2018, \aap, 616, A1

\bibitem[{{Cardelli} {et~al.}(1989){Cardelli}, {Clayton}, \&
  {Mathis}}]{1989ApJ...345..245C}
{Cardelli}, J.~A., {Clayton}, G.~C., \& {Mathis}, J.~S. 1989, \apj, 345, 245

\bibitem[{{Chabrier}(2003)}]{2003PASP..115..763C}
{Chabrier}, G. 2003, \pasp, 115, 763

\bibitem[{{Da Rio} \& {Robberto}(2012)}]{2012AJ....144..176D}
{Da Rio}, N., \& {Robberto}, M. 2012, \aj, 144, 176

\bibitem[{{Da Rio} {et~al.}(2014){Da Rio}, {Tan}, \&
  {Jaehnig}}]{2014ApJ...795...55D}
{Da Rio}, N., {Tan}, J.~C., \& {Jaehnig}, K. 2014, \apj, 795, 55

\bibitem[{{Da Rio} {et~al.}(2017){Da Rio}, {Tan}, {Covey}, {Cottaar}, {Foster},
  {Cullen}, {Tobin}, {Kim}, {Meyer}, {Nidever}, {Stassun}, {Chojnowski},
  {Flaherty}, {Majewski}, {Skrutskie}, {Zasowski}, \&
  {Pan}}]{2017ApJ...845..105D}
{Da Rio}, N., {Tan}, J.~C., {Covey}, K.~R., {et~al.} 2017, \apj, 845, 105

\bibitem[{{D'Orazi} {et~al.}(2009){D'Orazi}, {Randich}, {Flaccomio}, {Palla},
  {Sacco}, \& {Pallavicini}}]{2009A&A...501..973D}
{D'Orazi}, V., {Randich}, S., {Flaccomio}, E., {et~al.} 2009, \aap, 501, 973

\bibitem[{{Drass} {et~al.}(2016){Drass}, {Haas}, {Chini}, {Bayo}, {Hackstein},
  {Hoffmeister}, {Godoy}, \& {Vogt}}]{2016MNRAS.461.1734D}
{Drass}, H., {Haas}, M., {Chini}, R., {et~al.} 2016, \mnras, 461, 1734

\bibitem[{{Duquennoy} \& {Mayor}(1991)}]{1991A&A...248..485D}
{Duquennoy}, A., \& {Mayor}, M. 1991, \aap, 248, 485

\bibitem[{Foreman-Mackey(2016)}]{corner}
Foreman-Mackey, D. 2016, The Journal of Open Source Software, 24,
  doi:10.21105/joss.00024.
\newblock \url{http://dx.doi.org/10.5281/zenodo.45906}

\bibitem[{{Foreman-Mackey} {et~al.}(2013){Foreman-Mackey}, {Hogg}, {Lang}, \&
  {Goodman}}]{2013PASP..125..306F}
{Foreman-Mackey}, D., {Hogg}, D.~W., {Lang}, D., \& {Goodman}, J. 2013, \pasp,
  125, 306

\bibitem[{{Gennaro} {et~al.}(2015){Gennaro}, {Tchernyshyov}, {Brown}, \&
  {Gordon}}]{2015ApJ...808...45G}
{Gennaro}, M., {Tchernyshyov}, K., {Brown}, T.~M., \& {Gordon}, K.~D. 2015,
  \apj, 808, 45

\bibitem[{{Goodman} \& {Weare}(2010)}]{2010CAMCS...5...65G}
{Goodman}, J., \& {Weare}, J. 2010, Communications in Applied Mathematics and
  Computational Science, Vol.~5, No.~1, p.~65-80, 2010, 5, 65

\bibitem[{{Hillenbrand} \& {Carpenter}(2000)}]{2000ApJ...540..236H}
{Hillenbrand}, L.~A., \& {Carpenter}, J.~M. 2000, \apj, 540, 236

\bibitem[{{Hillenbrand} \& {Hartmann}(1998)}]{1998ApJ...492..540H}
{Hillenbrand}, L.~A., \& {Hartmann}, L.~W. 1998, \apj, 492, 540

\bibitem[{{Hopkins}(2013)}]{2013MNRAS.433..170H}
{Hopkins}, P.~F. 2013, \mnras, 433, 170

\bibitem[{{Jeffries} {et~al.}(2011){Jeffries}, {Littlefair}, {Naylor}, \&
  {Mayne}}]{2011MNRAS.418.1948J}
{Jeffries}, R.~D., {Littlefair}, S.~P., {Naylor}, T., \& {Mayne}, N.~J. 2011,
  \mnras, 418, 1948

\bibitem[{{Jerabkova} {et~al.}(2019){Jerabkova}, {Beccari}, {Boffin},
  {Petr-Gotzens}, {Manara}, {Prada Moroni}, {Tognelli}, \&
  {Degl'Innocenti}}]{2019A&A...627A..57J}
{Jerabkova}, T., {Beccari}, G., {Boffin}, H. M.~J., {et~al.} 2019, \aap, 627,
  A57

\bibitem[{{Koekemoer} {et~al.}(2011){Koekemoer}, {Faber}, {Ferguson}, {Grogin},
  {Kocevski}, {Koo}, {Lai}, {Lotz}, {Lucas}, {McGrath}, {Ogaz}, {Rajan},
  {Riess}, {Rodney}, {Strolger}, {Casertano}, {Castellano}, {Dahlen},
  {Dickinson}, {Dolch}, {Fontana}, {Giavalisco}, {Grazian}, {Guo}, {Hathi},
  {Huang}, {van der Wel}, {Yan}, {Acquaviva}, {Alexander}, {Almaini}, {Ashby},
  {Barden}, {Bell}, {Bournaud}, {Brown}, {Caputi}, {Cassata}, {Challis},
  {Chary}, {Cheung}, {Cirasuolo}, {Conselice}, {Roshan Cooray}, {Croton},
  {Daddi}, {Dav{\'e}}, {de Mello}, {de Ravel}, {Dekel}, {Donley}, {Dunlop},
  {Dutton}, {Elbaz}, {Fazio}, {Filippenko}, {Finkelstein}, {Frazer}, {Gardner},
  {Garnavich}, {Gawiser}, {Gruetzbauch}, {Hartley}, {H{\"a}ussler},
  {Herrington}, {Hopkins}, {Huang}, {Jha}, {Johnson}, {Kartaltepe},
  {Khostovan}, {Kirshner}, {Lani}, {Lee}, {Li}, {Madau}, {McCarthy},
  {McIntosh}, {McLure}, {McPartland}, {Mobasher}, {Moreira}, {Mortlock},
  {Moustakas}, {Mozena}, {Nandra}, {Newman}, {Nielsen}, {Niemi}, {Noeske},
  {Papovich}, {Pentericci}, {Pope}, {Primack}, {Ravindranath}, {Reddy},
  {Renzini}, {Rix}, {Robaina}, {Rosario}, {Rosati}, {Salimbeni}, {Scarlata},
  {Siana}, {Simard}, {Smidt}, {Snyder}, {Somerville}, {Spinrad}, {Straughn},
  {Telford}, {Teplitz}, {Trump}, {Vargas}, {Villforth}, {Wagner}, {Wandro},
  {Wechsler}, {Weiner}, {Wiklind}, {Wild}, {Wilson}, {Wuyts}, \&
  {Yun}}]{2011ApJS..197...36K}
{Koekemoer}, A.~M., {Faber}, S.~M., {Ferguson}, H.~C., {et~al.} 2011, \apjs,
  197, 36

\bibitem[{{Kounkel} {et~al.}(2017){Kounkel}, {Hartmann}, {Loinard},
  {Ortiz-Le{\'o}n}, {Mioduszewski}, {Rodr{\'{\i}}guez}, {Dzib}, {Torres},
  {Pech}, {Galli}, {Rivera}, {Boden}, {Evans}, {Brice{\~n}o}, \&
  {Tobin}}]{2017ApJ...834..142K}
{Kounkel}, M., {Hartmann}, L., {Loinard}, L., {et~al.} 2017, \apj, 834, 142

\bibitem[{{Kounkel} {et~al.}(2018){Kounkel}, {Covey}, {Su{\'a}rez},
  {Rom{\'a}n-Z{\'u}{\~n}iga}, {Hernandez}, {Stassun}, {Jaehnig}, {Feigelson},
  {Pe{\~n}a Ram{\'\i}rez}, {Roman-Lopes}, {Da Rio}, {Stringfellow}, {Kim},
  {Borissova}, {Fern{\'a}ndez-Trincado}, {Burgasser},
  {Garc{\'\i}a-Hern{\'a}ndez}, {Zamora}, {Pan}, \&
  {Nitschelm}}]{2018AJ....156...84K}
{Kounkel}, M., {Covey}, K., {Su{\'a}rez}, G., {et~al.} 2018, \aj, 156, 84

\bibitem[{{Kritsuk} {et~al.}(2017){Kritsuk}, {Ustyugov}, \&
  {Norman}}]{2017NJPh...19f5003K}
{Kritsuk}, A.~G., {Ustyugov}, S.~D., \& {Norman}, M.~L. 2017, New Journal of
  Physics, 19, 065003

\bibitem[{{Kroupa}(2001)}]{2001MNRAS.322..231K}
{Kroupa}, P. 2001, \mnras, 322, 231

\bibitem[{{Krumholz}(2011)}]{2011ApJ...743..110K}
{Krumholz}, M.~R. 2011, \apj, 743, 110

\bibitem[{{Krumholz}(2014)}]{2014PhR...539...49K}
---. 2014, \physrep, 539, 49

\bibitem[{{Kuhn} {et~al.}(2019){Kuhn}, {Hillenbrand}, {Sills}, {Feigelson}, \&
  {Getman}}]{2019ApJ...870...32K}
{Kuhn}, M.~A., {Hillenbrand}, L.~A., {Sills}, A., {Feigelson}, E.~D., \&
  {Getman}, K.~V. 2019, \apj, 870, 32

\bibitem[{{Larson}(1992)}]{1992MNRAS.256..641L}
{Larson}, R.~B. 1992, \mnras, 256, 641

\bibitem[{Lim(2018)}]{lim_pey_lian_2018_3247832}
Lim, P.~L. 2018, stsynphot, , , doi:10.5281/zenodo.3247832.
\newblock \url{https://doi.org/10.5281/zenodo.3247832}

\bibitem[{{Marsh} {et~al.}(2010){Marsh}, {Kirkpatrick}, \&
  {Plavchan}}]{2010ApJ...709L.158M}
{Marsh}, K.~A., {Kirkpatrick}, J.~D., \& {Plavchan}, P. 2010, \apjl, 709, L158

\bibitem[{{Muench} {et~al.}(2002){Muench}, {Lada}, {Lada}, \&
  {Alves}}]{2002ApJ...573..366M}
{Muench}, A.~A., {Lada}, E.~A., {Lada}, C.~J., \& {Alves}, J. 2002, \apj, 573,
  366

\bibitem[{{Pacifici} {et~al.}(2012){Pacifici}, {Charlot}, {Blaizot}, \&
  {Brinchmann}}]{2012MNRAS.421.2002P}
{Pacifici}, C., {Charlot}, S., {Blaizot}, J., \& {Brinchmann}, J. 2012, \mnras,
  421, 2002

\bibitem[{{Pacifici} {et~al.}(2016){Pacifici}, {Kassin}, {Weiner}, {Holden},
  {Gardner}, {Faber}, {Ferguson}, {Koo}, {Primack}, {Bell}, {Dekel}, {Gawiser},
  {Giavalisco}, {Rafelski}, {Simons}, {Barro}, {Croton}, {Dav{\'e}}, {Fontana},
  {Grogin}, {Koekemoer}, {Lee}, {Salmon}, {Somerville}, \&
  {Behroozi}}]{2016ApJ...832...79P}
{Pacifici}, C., {Kassin}, S.~A., {Weiner}, B.~J., {et~al.} 2016, \apj, 832, 79

\bibitem[{{Padoan} {et~al.}(2007){Padoan}, {Nordlund}, {Kritsuk}, {Norman}, \&
  {Li}}]{2007ApJ...661..972P}
{Padoan}, P., {Nordlund}, {\AA}., {Kritsuk}, A.~G., {Norman}, M.~L., \& {Li},
  P.~S. 2007, \apj, 661, 972

\bibitem[{{Raghavan} {et~al.}(2010){Raghavan}, {McAlister}, {Henry}, {Latham},
  {Marcy}, {Mason}, {Gies}, {White}, \& {ten Brummelaar}}]{2010ApJS..190....1R}
{Raghavan}, D., {McAlister}, H.~A., {Henry}, T.~J., {et~al.} 2010, \apjs, 190,
  1

\bibitem[{{Reggiani} {et~al.}(2011){Reggiani}, {Robberto}, {Da Rio}, {Meyer},
  {Soderblom}, \& {Ricci}}]{2011A&A...534A..83R}
{Reggiani}, M., {Robberto}, M., {Da Rio}, N., {et~al.} 2011, \aap, 534, A83

\bibitem[{{Robin} {et~al.}(2003){Robin}, {Reyl{\'e}}, {Derri{\`e}re}, \&
  {Picaud}}]{2003A&A...409..523R}
{Robin}, A.~C., {Reyl{\'e}}, C., {Derri{\`e}re}, S., \& {Picaud}, S. 2003,
  \aap, 409, 523

\bibitem[{{Salpeter}(1955)}]{1955ApJ...121..161S}
{Salpeter}, E.~E. 1955, \apj, 121, 161

\bibitem[{{Scandariato} {et~al.}(2011){Scandariato}, {Robberto}, {Pagano}, \&
  {Hillenbrand}}]{2011AA...533A..38S}
{Scandariato}, G., {Robberto}, M., {Pagano}, I., \& {Hillenbrand}, L.~A. 2011,
  \aap, 533, A38

\bibitem[{{Somers} \& {Pinsonneault}(2015)}]{2015ApJ...807..174S}
{Somers}, G., \& {Pinsonneault}, M.~H. 2015, \apj, 807, 174

\bibitem[{{Spitzer}(1969)}]{1969ApJ...158L.139S}
{Spitzer}, Lyman, J. 1969, \apjl, 158, L139

\bibitem[{{Stamatellos} \& {Whitworth}(2009)}]{2009MNRAS.392..413S}
{Stamatellos}, D., \& {Whitworth}, A.~P. 2009, \mnras, 392, 413

\bibitem[{{STScI development Team}(2018)}]{2018ascl.soft11001S}
{STScI development Team}. 2018, {synphot: Synthetic photometry using Astropy},
  Astrophysics Source Code Library, , , ascl:1811.001

\bibitem[{{Whitworth} {et~al.}(2007){Whitworth}, {Bate}, {Nordlund},
  {Reipurth}, \& {Zinnecker}}]{2007prpl.conf..459W}
{Whitworth}, A., {Bate}, M.~R., {Nordlund}, {\AA}., {Reipurth}, B., \&
  {Zinnecker}, H. 2007, Protostars and Planets V, 459

\bibitem[{{Zapatero Osorio} {et~al.}(2008){Zapatero Osorio}, {B{\'e}jar},
  {Bihain}, {Mart{\'{\i}}n}, {Rebolo}, {Vill{\'o}-P{\'e}rez},
  {D{\'{\i}}az-S{\'a}nchez}, {P{\'e}rez Garrido}, {Caballero}, {Henning},
  {Mundt}, {Barrado Y Navascu{\'e}s}, \& {Bailer-Jones}}]{2008A&A...477..895Z}
{Zapatero Osorio}, M.~R., {B{\'e}jar}, V.~J.~S., {Bihain}, G., {et~al.} 2008,
  \aap, 477, 895

\end{thebibliography}
\bibliographystyle{aasjournal}

\end{document}